\documentstyle[prl,aps,floats,epsf,epsfig]{revtex}

\begin{document}

\draft

\title{Exact Solutions to Special High Dimensional 
$O(n)$ Models, Dimensional Reductions, gauge redundancy,
and special Frustrated Spin and Orbital models}

\author{Zohar Nussinov}
\address{Institute Lorentz for Theoretical Physics, Leiden University\\
P.O.B. 9506, 2300 RA Leiden, The Netherlands}
\date{\today ; E-mail:zohar@lorentz.leidenuniv.nl}

\twocolumn[

\widetext
\begin{@twocolumnfalse}

\maketitle

\begin{abstract}

This work addresses models
(e.g. potential models 
of directed orbital
systems- the manganates)
in which an
effective reduction dimensionality
occurs as a result of a new
symmetry which is intermediate 
between that of global
and local gauge symmetry.
This path towards dimensional
reduction is examined in simple
$O(n)$ spin models and lattice 
gauge theories.
A high temperature expansion is employed to
map special anisotropic 
high dimensional models into 
lower dimensional variants.
We show that it is possible to have
an effective reduction in the dimension
without the need of compactifying some dimensions. 
These models are frustrated and 
display a symmetry intermediate between
local and global gauge symmetries. Some
solutions are presented. Our dimensional reductions
are a generlization of the trivial dimensional reduction
that occur in pure two 
dimensional gauge theories. It will be further
seen that the absence of a ``phase interference''
effect plays an important role in high dimensional 
problems. By identifying
another (``permutational'') symmetry
present in the large $n$ limit, we 
will further show 
how to generally map global
high dimensional spin 
systems onto a one dimensional
chain and discuss 
implications.

\end{abstract}

\vspace{0.5cm}

\narrowtext

\end{@twocolumnfalse}
]

\section{Introduction}

In this article, a high temperature expansion
is employed to examine mappings between problems
of various dimensionality.  By these
reductions we will exactly solve an 
anisotropic two dimensional XY and 
some other models.  It will be 
seen that the absence of a ``phase interference''
effect plays an important role in high dimensional 
problems; this observation 
will allow us to examine various 
high dimensional models by examining newly 
constructed one dimensional systems.
We will show how an
effective reduction dimensionality
might occur as a result of an enhanced 
symmetry which lies midway
between a full blown local
symmetry and the 
much more restricted global
symmetry. Such a symmetry
might lead to 
an effective reduction in the dimension
{\em without the need of actually compactifying } any
dimensions.
By further identifying
a permutational symmetry present
in the large $n$ limit, we 
will further show 
how to generally map global
high dimensional spin 
systems onto a one dimensional
chain and discuss implications.

In section \ref{a},  We are going to start off by discussing 
a momentum space high temperature 
expansion. Next, by analytically continuing
the momenta and making them complex 
(section \ref{exact}), quaternion (section
\ref{high}), and of a general matrix form we will
be able to exactly solve a variety 
of two, three, and four dimensional
Ising and $O(n)$ models. Many of 
these models are artificial 
and correspond to non hermitian 
anisotropic long range 
interactions. Some of the resulting
models will be short ranged and of real 
hermitian form. We will exactly solve
some special simple two dimensional 
anisotropic XY models. These models
may serve as simple caricatures
for directed orbital models
for transition metal oxides (e.g.
the manganates). 
These models possess a symmetry
which is intermediate 
between that of
global and local
theories. Similar models
also occur naturally in the 
low energy regime of
frustrated pyrochlore
and checkerboard 
antiferromagnets.

We will then extend these techniques to the 
quantum case and discuss 
``2+1 dimensional Bethe 
Ansatz'' solutions in 
section(\ref{Bethe})
where we will also
solve a special two dimensional
anisotropic XY model.

Next, we show in Section(\ref{GAUGE})
that the trivial dimensional
reduction of a 1+1 dimensional
gauge theory to a one dimensional
spin chain can be thought of
as a trivial example of
our general mode for
dimensional reductions.

After solving these models, in subsection(\ref{comm})
we will apply the high
temperature expansion to
reduce the three dimensional
nearest neighbor Ising ferromagnet
into an illusory {\em formal} solvable 
expression written in a determinant
notation. In this method, the 
high temperature coefficient to each order 
is arrived at by looking a {\em single linear relation}
originating from a matrix of 
(super exponentially) increasing order.

In subsection(\ref{incomm}),  we will  
map the nearest neighbor $d$ dimensional 
Ising models to a one dimensional model
whose spectrum is the sum of $d$ 
mutually incommensurate tight binding
like spectra.
This in turn will enable us
to map the $d$ dimensional Ising
model onto a
a single spin 1/2 quantum particle
in 0+1 dimensions.

In section(\ref{permutational}), 
we will discuss a ``permutational''
symmetry present in the $O(n \rightarrow
\infty)$ models but violated to 
${\cal{O}}(\beta^{4})$ in finite 
$n$ problem. This will allow 
us to map any large $n$ problem onto 
a one dimensional lattice system. 

One of the main motifs 
of this article is the 
 absence or presence of
``momentum interference''
effects in high and
low dimensional problems 
respectively.
Another recurring
motif is that of 
a novel symmetry
which interpolates
between global
and local symmetry
and which effectively
lowers the dimensionality
of the problem.

\bigskip

\section{A momentum space high temperature 
expansion}
\label{a}

We will for the most part examine simple classical 
spin models of the type

\begin{eqnarray}
H = \frac{1}{2} \sum_{\vec{x},\vec{y}}
\hat{V}(\vec{x},\vec{y})[\vec{S}(\vec{x}) 
\cdot \vec{S}(\vec{y})]. 
\label{te}
\end{eqnarray} 

Here, the sites $\vec{x}$ and $\vec{y}$
lie on a (generally hypercubic)  
lattice of size $N$. The spins $\{ S(\vec{x}) \}$
are normalized and have $n$ components. The kernel
$\hat{V}(\vec{x},\vec{y})$ is translation invariant.
The $n=1$ case simply corresponds to the 
Ising model.

\bigskip

Later on, when we will discuss quaternion models
we will replace 
the scalar product by a slightly more
complicated product. 

\bigskip

The Hamiltonian in the (non-symmetrical)
Fourier basis is diagonal 
($f(\vec{k}) = \sum _{\vec{x}} 
F(\vec{x}) e^{-i \vec{k} \cdot \vec{x}};$
\bigskip
$ ~ F(\vec{x}) = \frac{1}{N} \sum_{\vec{k}} 
f(\vec{k}) e^{i \vec{k} \cdot \vec{x}}$) and reads 

\begin{eqnarray}
H =  \frac{1}{2N} \sum_{\vec{k}} \hat{v}(\vec{k}) |\vec{s}(\vec{k})|^{2},
\end{eqnarray}

where $\hat{v}(\vec{k})$ and $\vec{s}(\vec{k})$ are the Fourier
transforms of $\hat{V}(\vec{x})$ and $\vec{S}(\vec{x})$
respectively.

For simplicity, we will set the lattice constant to unity-
i.e.  on a hypercubic lattice (of side $L$) with
periodic boundary conditions (which we will
assume throughout) the wave-vector
components $k_{l} = \frac{2 \pi r_{l}}{L}$
where $r_{l}$ is an integer (and the real space 
coordinates  $x_{l}$ are integers).

In the up and coming $V \equiv \beta \hat{V}$. For 
an invertible $\hat{V}(\vec{x},\vec{y})$,
and an inverse temperature $0<\beta<\infty$ (via a Hubbard-Stratonovich 
transformation \cite{Hubbard-Statonovich}, see also Appendix A), the 
partition function:
\begin{eqnarray}
  Z=Tr \exp\{-\frac{1}{2}
  \sum_{\vec{x},\vec{y}}V(\vec{x},\vec{y})S(\vec{x})S(\vec{y})\}
\nonumber 
\\ =
  \sqrt{\det(-\frac{2V}{\pi})}\int\prod_{\vec{x}} d\eta(\vec{x})
  \exp[-\tilde{H}^{dual}\{\eta(\vec{x})\}]
\end{eqnarray}
where
\begin{eqnarray}
  \tilde{H}^{dual}[\{\eta(\vec{x})\}] = -\frac{1}{2} \sum_{\vec{x},\vec{y}}
  \eta(\vec{x})V^{-1}(\vec{x},\vec{y})\eta(\vec{y})\nonumber
\\ -\sum_{\vec{x}}\ln[\cosh
  \eta(\vec{x})]
\label{CoSh-}
\end{eqnarray}
and analogously for the $O(n>1)$ model
\begin{eqnarray}
  \tilde{H}^{dual}\{\vec{\eta}(\vec{x})\}= -\frac{1}{2} \sum_{\vec{x},\vec{y}}
  V^{-1}(\vec{x},\vec{y})\vec{\eta}(\vec{x})\cdot \vec{\eta}(\vec{y})
\nonumber
\\ 
  -\sum_{\vec{x}}\ln[
  (\frac{2}{|\vec{\eta}(\vec{x})|})^{n/2-1}~
  I_{n/2-1}(|\vec{\eta}(\vec{x})|)~]
\label{Bes}
\end{eqnarray}
with  $I_{n/2-1}(z)$ a Bessel function.

The kernel within the dual Hamiltonian $\tilde{H}^{dual}$ 
is the inverse of the kernel $V$ appearing
in the original Hamiltonian. It is Coulomb 
like \cite{Cool} for the nearest neighbor kernel  
$\hat{V}(\vec{x},\vec{y}) = - \delta_{|\vec{x}-\vec{y}|,1}$.

The $\cosh \eta$ and Bessel function terms
appearing in Eqs.(\ref{CoSh-}) and (\ref{Bes}) 
may be viewed as terms of constraint
securing the normalization
of $\vec{S}(\vec{x})$ at 
every site $\vec{x}$.
The rotational invariance
of the scalar product in
Eqn.(\ref{te}) is manifested
by the rotational invariance 
($|\vec{\eta}|$ dependence)
of the Bessel function 
terms.

If a magnetic field $h(\vec{x})$ were
applied, the argument of the $\cosh$ 
would be replaced $\cosh(\eta+ h)$ with 
a similar occurrence ($|\vec{\eta}| \to |\vec{\eta}+ \vec{h}|$)
for the Bessel function appearing for the 
$O(n>1)$ models.

The $q-$state Potts model 
can be viewed as a spin
model in which the $q$ possible
polarizations of the spins
lie at the vertices of a $(q-1)$ 
dimensional tetrahedron- in this manner
the scalar product amongst
any two non identical spins
is $\cos^{-1}(-1/(q-1))$. Employing this
representation, the sum $\sum_{\vec{x}} \ln \cosh \eta(\vec{x})$
may be replaced by $\sum_{\vec{x}} \ln \sum_{i=1}^{q}
\exp[\vec{S}_{i} \cdot \vec{\eta}]$
where the sum is over the $q$ polarizations
of the $(q-1)$ dimensional spin 
$\vec{S}_{i}$.

For a $q$ state clock model, the Bessel functions
are similarly replaced by $\sum_{\vec{x}} 
\ln \sum_{i=1}^{q} \exp[\vec{S}_{i} \cdot 
\vec{\eta}]$. For the 4-state clock model 
which we will consider later on
this is $\sum_{x} \ln [2(\cosh \eta_{1} + 
\cosh \eta_{2})]$. We will 
introduce, in section(\ref{high}),
in the context of the directed
electronic orbitals appearing in 
the manganates, ``directed 
clock-models''. These models
share the same generating
function  $\sum_{\vec{x}} 
\ln \sum_{i=1}^{q} \exp[\vec{S}_{i} \cdot 
\vec{\eta}]$ albeit with a polarization 
($\alpha$) dependent kernel $\hat{V}_{\alpha \beta}
(\vec{x},\vec{y})$.

The partition function of the 
Ising spins reads
\begin{eqnarray}
Z=  \sqrt{\det(-\frac{2V}{\pi})}\int D \eta e^{- \beta H} \nonumber
\\ =  \sqrt{\det(-\frac{2V}{\pi})}\int \prod_{\vec{x}} d\eta(\vec{x})
\nonumber
\\ \exp \Big[-\frac{1}{2} \sum_{\vec{x},\vec{y}} \eta(\vec{x})
V^{-1}(\vec{x},\vec{y}) \eta(\vec{y}) \Big] \nonumber
\\ \prod_{\vec{x}} \cosh \eta(\vec{x}).
\end{eqnarray}
For a translationally invariant 
interaction $V(\vec{x},\vec{y}) = V(\vec{x}-\vec{y})$, 
this may be written in momentum space
\begin{eqnarray}
Z=  \sqrt{\det(-\frac{2V}{\pi})}\int 
D \eta \exp \Big[-\frac{1}{2N} \nonumber
\\ \sum_{\vec{k}} v^{-1}(\vec{k})
\eta(\vec{k}) \eta(-\vec{k}) \Big] \nonumber
\\ \prod_{\vec{x}} \sum_{m=0}^{\infty} \frac{\eta^{2m}(\vec{x})}{(2m)!},
\end{eqnarray} 
where $v(\vec{k})$ is the Fourier transform
\begin{eqnarray}
v(\vec{k}) = \sum_{\vec{x}} V(\vec{x}-\vec{y}) e^{ i \vec{k} \cdot
\vec{x}} \equiv \beta \hat{v}(\vec{k}).
\end{eqnarray}
where the sum is over all lattice
sites $\vec{x}$. 
Thus the partition function is , trivially, 
\begin{eqnarray} 
Z =  {\cal{N}} \sum_{m_{1},...m_{N}} \langle 
\prod_{i=1}^{N} \frac{1}{(2 m_{i})!}
 \eta^{2m_{i}}(\vec{x}_{i})\rangle_{0} 
\label{non-asym}
\end{eqnarray}
where $\langle ~ ~ \rangle_{0}$ denotes an average 
with respect to the unperturbed Gaussian weight 
\begin{equation}
\exp \Big[-\frac{1}{2N} \sum_{\vec{k}} v^{-1}(\vec{k})
\eta(\vec{k}) \eta(-\vec{k}) \Big],
\end{equation}
and ${\cal{N}}$ is 
a normalization constant.

Each contraction $\langle \eta(\vec{x}) \eta(\vec{y}) \rangle$
(or $\langle \eta(\vec{k}) \eta(-\vec{k}) \rangle$)
leads to a factor of  $V(\vec{x},\vec{y}) = \beta
\hat{V}(\vec{x},\vec{y})$  
(or to $\beta \hat{v}(\vec{k})/N$ in momentum 
space). Thus the resultant series 
is an expansion in the
inverse temperature $\beta$. In the up and coming we will 
focus attention on the momentum space formulation
of this series. All the momentum space algebra
presented above is only a slight modification 
to the well known high temperature expansions usually
generated by the Hubbard Stratonovich transformation 
directly applied to the real space representation
of the fields $\eta(\vec{x})$ \cite{Wortis,Polyakov}. 
It is due to the naivete of the author that 
such a redundant momentum space formulation
was rederived in the first place. However, 
as we will see, in momentum space some properties 
of the series become much more transparent. To 
make $V$ invertible and the expansion convergent, 
we will shift $\hat{v}(\vec{k})$ by a constant 
\begin{equation}
\hat{v}(\vec{k}) \rightarrow \hat{v}(\vec{k}) + A
\end{equation}
with $A = const> 2d$
such that $\hat{v}(\vec{k})$ is strictly positive. 
In real space such a constant shift amounts
to a trivial shift in the on site interaction 
(or chemical potential)
\begin{eqnarray}
\hat{V}(\vec{x},\vec{y}) \rightarrow \hat{V}(\vec{x},\vec{y}) +A.
\delta_{\vec{x},\vec{y}}
\end{eqnarray}
For asymptotically large $|\eta|$ the nontrivial 
\begin{eqnarray}
\tilde{H}^{dual}_{1} \equiv - \sum_{\vec{x}} \ln |\cosh \eta(\vec{x})|
\end{eqnarray}
is linear in $|\eta|$. The Gaussian generating
\begin{eqnarray}
\tilde{H}^{dual}_{0} = - \frac{1}{2} \sum_{\vec{x},\vec{y}}
  \eta(\vec{x})V^{-1}(\vec{x},\vec{y})\eta(\vec{y})
\end{eqnarray}
is a positive quadratic
form and dominates at large 
$\eta$.  Note that the convergence
of this series in Eqn.(\ref{non-asym}) is better
than that appearing in the 
canonical field theories
where the perturbing $H_{1}$ dominates
over $H_{0}$ for 
large fields and, as a consequence, 
all canonical expansions are blatantly
asymptotic \cite{Blat}.

So far all of this
has been very general.
Now let us consider the
$d$ dimensional nearest 
neighbor Ising model. 
To avoid carrying minus signs around, we will consider 
 an Ising antiferromagnet whose 
partition function is, classically, identically 
equal to that of a ferromagnet. The exchange 
constant $J$ will be set to unity. The corresponding momentum space
kernel 
\begin{equation}
\hat{v}(\vec{k}) =
A + \sum_{l=1}^{d} [e^{i k_{l}} + e^{-i k_{l}}].
\label{d-dim}
\end{equation}

The resulting high temperature 
series ($\ln Z = - N \sum_{p} f_{p} \beta^{p}$) 
must be identical to the
one derived by conventional
real space methods which
is convergent for all $\beta <\beta_{c}$ (the inverse 
critical temperature): we retiterate that this convergence
proves that the series we obtain is not asymptotic
and free from many of the pathologies
of common diagrammatic expansions \cite{Blat}.

As the reader might guess, the uniform shift $A$ 
only leads to a trivial change in $Z$
(i.e. to a multiplication by the constant 
$\exp[- \beta A N]$).
In the forthcoming we will set $A=0$.

Owing to the 
relation
\begin{equation}
\int_{-\pi}^{\pi} dk ~e^{ikN^{\prime}} = 2 \pi \delta_{N^{\prime},0}
\label{int}
\end{equation}
all loop integrals are
trivial.

All stated here and below also 
holds exactly for finite size 
(of sides $L>1$) systems
with periodic boundary conditions.
Here 
\begin{eqnarray}
\sum_{k_{n}} \exp[i k_{n} N^{\prime}] = 
L \delta_{N^{\prime},0}
\label{lat-summ}
\end{eqnarray}
 replaces 
Eqn.(\ref{int}).
In Eqn.(\ref{lat-summ})
the sum is over all $k_{n} = \frac{2 \pi n}{L}$ 
with $-\pi  < k_{n} \le \pi$.
Note, however, that Eqn.(\ref{lat-summ}) is far
more general and holds for 
a large class of finite 
size systems.
\cite{tsitsi}.

In the thermodynamic limit, a simple integral
is of the type 

\begin{eqnarray} 
\int_{\sum_{i=1}^{M} \vec{k}_{i}=0}
\prod_{i=1}^{M} \frac{d^{d}k_{i}}{(2 \pi)^{d}} \prod_{i} 
\hat{v}^{n}(\vec{k}_{i}) \nonumber
\\ = \sum_{n=\sum_{l=1}^{d} (\gamma_{l}+\delta_{l})} [ \frac{n!}{\prod_{l=1}^{d} \gamma_{l}! \delta _{l}!}]^{M}
\end{eqnarray}
(when $n=1$ this integral is $2d$ etc).
A related integral reads 
\begin{eqnarray}
\int \frac{d^{d}k}{(2 \pi)^{d}}
\prod_{\sum_{i=1}^{M} \vec{k}_{i}=0}~~
[\hat{v}(\vec{k}_{i})]^{n_{i}} \nonumber
\\ = \sum_{\gamma_{l}^{i}-\delta_{l}^{i}
= m_{i}^{l},~\sum_{i=1}^{M} m_{i}^{l} =0,~ 
\sum_{l=1}^{d} \gamma_{l}^{i}+\delta_{l}^{i} = n_{i}}
\prod_{i=1}^{M} [\frac{n_{i}!}{\prod_{l=1}^{d} \gamma_{l}^{i}! \delta _{l}^{i}!}]
\end{eqnarray}
where in the last sum $m_{j<n}^{l} \neq m_{i <M}^{l} = -m_{i=M}^{l}$
etc. 
Note that the high temperature expansion 
becomes simpler if instead of symmetrizing
the interaction (each bond being counted twice-
once by each of the two interacting spins) 
one considers 
\begin{equation}
\hat{v}(\vec{k}) = 2 \sum_{l=1}^{d} \exp[i k_{l}].  
\end{equation}
Here each spin interacts with only $d$ (and not $2d$) of
its neighbors separated from it 
by one positive distance along
the d axes $\hat{e}_{l=1}^{d}$. 
The factor of 2 originates as each 
bond is now counted only once
and therefore the corresponding bond 
strength is doubled.
The partition function Z is 
trivially unchanged. 

In the most general diagram
the propagator momenta $\{ \vec{q}_{a} \}$
($1 \le a \le$ number of propagators)
are linear combinations of the 
independent loop momenta $\{ \vec{k}_{b} \}_{b=1}^{\mbox{loops}}$
\begin{eqnarray}
\vec{q}_{a} = M_{ab} \vec{k}_{b}
\end{eqnarray}
with integer coefficients $M_{ab}$.

Symmetry factors aside,
the value of a given 
diagram reads
\begin{eqnarray}
\int \prod_{b} \frac{d^{d}k_{b}}{(2 \pi)^{d}} \prod_{a} v(\sum M_{ab}
k_{b})  \nonumber
\\ = \int \prod_{b} \frac{d^{d}k_{b}}{(2 \pi)^{d}} \prod_{a} (2 \beta)
\sum_{l=1}^{d}  \exp[i \sum_{b} M_{ab} k_{b}^{l}] 
\nonumber
\\ =  (2 \beta)^{\mbox{propagators}} \nonumber
\\ \times  \Big( \int \prod_{b=1}^{\mbox{loops}} \frac{d^{d}k_{b}}{(2 \pi)^{d}}
\sum_{l=1}^{d}  \exp[i \sum_{a,b} M_{ab} k_{b}^{l}] \Big)
\label{loop}
\end{eqnarray}
- i.e. upon expansion of the outer sum, 
just a product of individual
loop integrals of the type encountered
in Eqn.(\ref{int}) for each $k^{l}_{b}$.

The latter integral in $d=1$ reads
\begin{eqnarray} 
I = \prod_{b=1}^{\mbox{loops}} \delta(\sum_{a=1}^{\mbox{propagators}} 
M_{ab},0)
\end{eqnarray}
with a Kronecker delta.
The situation in $d >1$ becomes far richer. 
One may partition the integral into all
possible subproducts of momenta corresponding
to different spatial components $l$ satisfying
the delta function constraints.
In a given diagram not all propagators need
to correspond to the same spatial 
component $k^{l}$. The combinatorics of
``in how many ways loops may be chosen
to satisfy the latter delta
function constraints?'' boils 
down to the standard counting 
of closed loops in real 
space.  Explicitly, in general $d$,
\begin{eqnarray}
 I = \sum_{\mbox{all partitions of $\{q_{a}\}$ into d sets
${\cal{C}}_{l}$}} ~ \prod_{l=1}^{\mbox{d}} \nonumber
\\
\prod_{b=1}^{\mbox{loops}}
\delta(\sum_{a \in {\cal{C}}_{l}} M_{ab}, 0).
\label{color}
\end{eqnarray}

Note that entirely identical
expressions would be reached
for $O(n>1)$ models- the diagrams
to be drawn and their integral
expressions s are exactly the same
as for the Ising system. The sole 
difference is that in the $O(n>1)$ case,
the vertices of various orders 
(given by the Taylor expansion 
of the Bessel functions) will
have different weights.

We may imagine coloring, for each individual term
in Eqn.(\ref{color}), the different sets  ${\cal{C}}_{l}$ 
of propagator lines $\{q_{a}^{l}\}$ corresponding to different 
dimensions by different colors to produce a graph colored 
with $d$ colors. It is amusing to note that, by the four-color 
theorem, the most general topology 
of planar diagrams is present
at $d=4$ dimensions.

Some simple technical details of 
lattice perturbation theory
vis a vis the standard continuum
theories are presented in 
\cite{N}.

\section{An exact solution
to a special three dimensional Ising
model}
\label{exact}

If the spin spin interaction on a cubic lattice 
has a non hermitian kernel of
the form 
\begin{eqnarray}
\hat{V}(x_{1},x_{2},x_{3}) = - \delta_{x_{2},0}
\delta_{x_{3},0}  (\delta_{x_{1},1}+ \delta_{x_{1},-1}) 
\nonumber
\\ - \frac{1}{\pi} (-1)^{x_{3}} \frac{\sinh \pi \lambda}{\lambda^{2}+ x_{3}^{2}}
 \delta_{x_{1},0} \Big[\lambda  (\delta_{x_{2},1}+ \delta_{x_{2},-1}) \nonumber
\\ + i x_{3} ( \delta_{x_{2},1}- \delta_{x_{2},-1}) 
 \Big]
\end{eqnarray}
then, for all real $\lambda$,
the Helmholtz free energy per spin
$f(h=0,\beta > \ln (1+ \sqrt{2})/2)$ is
\begin{eqnarray}
\beta f = - \ln( 2 \cosh 2 \beta) -
\nonumber
\\ 
\frac{1}{2 \pi} \int_{0}^{\pi} d \phi \ln \frac{1}{2}
(1+ \sqrt{1 - \kappa^{2} \sin^{2} \phi})
\end{eqnarray}
where
\begin{eqnarray}
\kappa \equiv \frac{2 \sinh 2 \beta}{\cosh^{2} 2 \beta}
\end{eqnarray}
and the internal energy per spin
\begin{eqnarray}
u=  - \coth 2 \beta  \Big[ 1+ \frac{2}{\pi}
\kappa^{\prime} K_{1}(\kappa) \Big]
\end{eqnarray}
where $K_{1}(\kappa)$ is a complete elliptic
integral of the first kind
\begin{eqnarray}
K_{1}(\kappa) \equiv \int_{0}^{\pi/2} \frac{d \phi}{\sqrt{1
-\kappa^{2} \sin^{2} \phi}}
\end{eqnarray}
and 
\begin{equation}
\kappa^{\prime} \equiv 2 \tanh^{2} (2 \beta)  -1.
\end{equation}

The proof is quite simple. The
expressions presented are
the corresponding intensive quantities for 
the two dimensional nearest neighbor Ising model\cite{Onsager}.

Let us start by writing down the 
diagrammatic expansion (absolutely convergent 
for $T> T_{c}$) for
the two dimensional Ising
model (where $T_{c} = 2/ \ln (1+ \sqrt{2})$ 
and set
\begin{eqnarray}
\hat{v}(q_{1},q_{2}) \rightarrow \hat{v}(q_{1},q_{2}+ i \lambda q_{3})
\end{eqnarray}
in all integrals. This effects
\begin{eqnarray}
\int_{-\pi}^{\pi} dk_{2} e^{i k_{2}N^{loop}} \rightarrow \int_{-\pi}^{\pi}
dk_{2}
e^{ik_{2}N^{loop}} \int_{-\pi}^{\pi} dk_{3} e^{-\lambda k_{3} N^{loop}}
\end{eqnarray}
for all individual loop integrals.
The integrals are nonvanishing only 
if $\{N_{b}=0 \}$ for all loops $b$. 
When $N^{loop}=0$ for a given loop
then the second ($k_{3}$) integration
leads to a trivial multiplicative
constant (by one). In the original
summation 
\begin{eqnarray}
\sum_{\vec{k}_{b}} \rightarrow \frac{L^{d}}{(2 \pi)^{d}} \int_{-\pi}^{\pi} ... 
\int_{-\pi}^{\pi} d^{d}k_{b}
\end{eqnarray}
and thus  an additional 
factor of $L$ is introduced for each independent connected
diagram. Thus the free energy per unit volume 
for a system with the momentum
space kernel $\hat{v}(k_{1},k_{2}+i \lambda k_{3})$
is the same as for the two dimensional
nearest neighbor ferromagnet with
the kernel $\hat{v}(k_{1},k_{2})$.
Fourier transforming 
 $\hat{v}(k_{1},k_{2}+i \lambda k_{3})$
and symmetrizing $[V(\vec{x}) + V(-\vec{x})]/2 \rightarrow V(\vec{x})$
one finds the complex real space
kernel $\hat{V}(x_{1},x_{2},x_{3})$
presented above \cite{Gibbs}

\bigskip

For the more general 
\begin{eqnarray}
\hat{V}(x_{1},x_{2},x_{3}) = -\delta_{x_{2},0} \delta_{x_{3},0}
(\delta_{x_{1},1} + \delta_{x_{1},-1}) \nonumber
\\ 
- {\cal{J}} \delta_{x_{1},0}\Big[\frac{(-1)^{x_{3}}}{\lambda^{2}+ x_{3}^{2}}\Big] 
\{ (\delta_{x_{2},1}+ \delta_{x_{2},-1})  \nonumber
\\ + i
(\delta_{x_{2},-1}-\delta_{x_{2},1}) \frac{x_{3}}{\lambda} \}
\label{Neel}
\end{eqnarray}
we may define 
\begin{eqnarray}
r \equiv \frac{\pi {\cal{J}}}{2 \lambda ~\sinh \pi \lambda}
\end{eqnarray}
to write the free energy
as the free energy of an 
anisotropic two dimensional 
nearest neighbor ferromagnet
with the kernel
\begin{eqnarray}
\hat{V}(x_{1},x_{2})= -\delta_{x_{2},0} (\delta_{x_{1},1} +
\delta_{x_{1},-1} \nonumber
\\ - r \delta_{x_{1},0}(\delta_{x_{2},1} + \delta_{x_{2},-1})
\end{eqnarray}
i.e. a nearest neighbor Ising ferromagnet having the 
ratio of the exchange constants along the $x_{1}$ and $x_{2}$
axes equal to $r= J_{2}/J_{1}$ (throughout we set $J_{1}$ to unity).
Introducing
\begin{eqnarray}
u \equiv \frac{1}{\sinh 2 \beta \sinh 2 \beta r}
\end{eqnarray}
and defining 
\begin{eqnarray}
F(\theta) \equiv \ln \{2 [\cosh 2 \beta \cosh 2 \beta r \nonumber
\\ + u^{-1} (1- 2 u
\cos \theta + u^{2})^{1/2}]\},
\end{eqnarray}
we may write the free energy density
as
\begin{equation}
\beta f = - \frac{1}{2 \pi} \int_{0}^{\pi} F(\theta)~ d \theta.
\end{equation}

\bigskip

\bigskip

Note that the same result 
applies for the ``ferromagnetic''

\begin{eqnarray}
\hat{V}(x_{1},x_{2},x_{3}) = -\delta_{x_{2},0} \delta_{x_{3},0}
(\delta_{x_{1},1} + \delta_{x_{1},-1}) \nonumber
\\ 
- {\cal{J}} \delta_{x_{1},0}\Big[\frac{1}{\lambda^{2}+ x_{3}^{2}}\Big] 
\{ (\delta_{x_{2},1}+ \delta_{x_{2},-1})  \nonumber
\\ + i
(\delta_{x_{2},-1}-\delta_{x_{2},1}) \frac{x_{3}}{\lambda} \}.
\label{ferro}
\end{eqnarray}
The proof is simple- the partition function
is identically the same if evaluated with the
kernel in Eqn.(\ref{ferro}) instead
of that  in Eqn.(\ref{Neel})
if the spins are flipped
\begin{eqnarray}
S(\vec{x}) \rightarrow (-1)^{x_{3}} S(\vec{x}).
\end{eqnarray}

\bigskip

\bigskip

What about correlation functions?
The equivalence of the partition
functions $Z[\beta,h(x_{1},x_{2})]$
implies that in the 
planar directions
the correlations
are the same as
for two dimensional ferromagnet.

Changing the single momentum
coordinate
\begin{eqnarray}
v(k_{1},k_{2}) \rightarrow v(k_{1},k_{2}+ i \lambda k_{3}) \nonumber
\end{eqnarray}
effects 
\begin{eqnarray}
G(k_{1},k_{2}) \rightarrow G(k_{1},k_{2}+ i \lambda k_{3})
\label{cor}
\end{eqnarray}
in the momentum space correlation function.
The algebra is identically the same.

The connected spatial correlations (for $T \neq T_{c}$) 
are exponentially  damped 
along the $x_{2}$ axis 
as they are in the usual two dimensional 
nearest neighbor ferromagnet (see \cite{Wu,Mc} 
for the two dimensional correlation
function).

Of course, all of this can also be extended
to one-dimensional like spin systems.

The two dimensional spin-spin kernel
\begin{eqnarray}
\hat{V}(x_{1},x_{2}) =   - \frac{1}{\pi} (-1)^{x_{2}} \frac{\sinh \pi \lambda}{\lambda^{2}+ x_{2}^{2}}
 \delta_{x_{1},0} \Big[\lambda  (\delta_{x_{2},1}+ \delta_{x_{2},-1}) \nonumber
\\ + i x_{2} ( \delta_{x_{2},1}- \delta_{x_{2},-1})] 
 \Big]
\label{special}
\end{eqnarray}
leads to one dimensional behavior
with oscillations 
along the $x_{2}$ axis.
More precisely, the  momentum space
correlator for any $O(n)$ system in
one dimension reads
\begin{eqnarray}
G(k) =  \frac{e^{r}}{e^{r}-e^{ik}} + \frac{1}{e^{ik+r}-1}.
\end{eqnarray}
For the Ising ($n=1$) system $r \equiv - [\ln \tanh \beta]$.
One may set
$k = k_{1} + i \lambda k_{2}$ and 
compute the inverse 
Fourier transform.
The correlations along 
$x_{2}$ are oscillatory
with temperature dependent 
wavevectors.  In other
words, the effective correlation length
along $x_{2}$ is infinite.

If the two dimensional spin-spin interactions
in Eqn.(\ref{special}) are augmented by an additional 
on-site magnetic field $h$ then the two dimensional 
partition function reads 
\begin{equation}
Z = \lambda_{+}^{N}+ \lambda_{-}^{N}
\end{equation}
where, as usual, 
\begin{eqnarray}
\lambda_{\pm} = e^{\beta} \cosh \beta h \pm \sqrt{e^{2 \beta}
\sinh^{2}\beta h + e^{-2 \beta}}.
\end{eqnarray}

That ``complexifying'' the coordinates 
should not change the physics is intuitively
obvious: if one shifts $k_{1} \rightarrow k_{1} + \lambda k_{2} \equiv
k^{\prime}_{1}$
with real $\lambda$ in the one dimensional 
$\hat{v}(k_{1})$ then the resulting 
kernel $\hat{v}(k^{\prime}_{1})$ plainly describes 
a stack on chains parallel to $(1,\lambda)$; the 
spins interact along the chain direction yet
the chains do not interact amongst themselves. 
The free energy density should be identically
equal to that of a one dimensional 
system- no possible dependence on 
$\lambda$ can occur. All that we have 
done in the above is allow
$\lambda$ to become complex.

In 
\begin{eqnarray}
\hat{v}(k_{1},k_{2}) \rightarrow \hat{v}(k_{1},k_{2}+ \lambda k_{3})
\end{eqnarray}
the correlation functions along
a direction orthogonal to the 
direction $(1,\lambda)$ in the $(x_{2},x_{3})$
plane vanish. If $\lambda$ is complex 
then there are no correlations along an
orthogonal direction in the 
``complex'' space.

Formally, all this stems from the trivial fact
that the measure $dk_{1} dk_{2} = const (dk dk^{*})$,
whereas $\hat{v}(k)$ depends only on the 
single complex coordinate $k$.

Other trivial generalizations
are possible. For example, we 
may make both $k_{1}$ and $k_{2}$
complex in $\hat{v}(k_{1},k_{2})$
to generate a four dimensional spin kernel
\begin{eqnarray}
\hat{V}(x_{1},x_{2}.x_{3},x_{4}) \nonumber
\\=
 - \frac{1}{\pi} \Big( (-1)^{x_{2}} \frac{\sinh \pi \lambda_{1}}{\lambda_{1}^{2}+ x_{2}^{2}}
 \delta_{x_{1},0} \Big[\lambda_{1}  (\delta_{x_{2},1}+ \delta_{x_{2},-1}) \nonumber
\\ + i x_{3} ( \delta_{x_{2},1}- \delta_{x_{2},-1})] 
 \Big] \nonumber
\\ +  (-1)^{x_{4}} \frac{\sinh \pi \lambda_{2}}{\lambda_{2}^{2}+ x_{4}^{2}}
 \delta_{x_{3},0} \Big[\lambda_{2}  (\delta_{x_{4},1}+
\delta_{x_{4},-1}) 
\nonumber \\ + i x_{4} ( \delta_{x_{3},1}- \delta_{x_{3},-1})] 
 \Big]  \Big)
\end{eqnarray}

which leads to the canonical (nearest neighbor) 
two dimensional behavior for arbitrary 
$\lambda_{1}$ and $\lambda_{2}$.

All of this
need not be restricted
to cubic lattice models. We may also
analytically continue $\hat{v}(\vec{k})$
of the triangular lattice etc.

For the triangular antiferromagnet/ferromagnet
\begin{eqnarray}
v(\vec{k}) = \pm 2[\cos k_{1} + \cos (\frac{{k}_{1}}{2} +
\frac{\sqrt{3}}{2} k_{2}) \nonumber
\\ + \cos(-\frac{k_{1}}{2}+\frac{\sqrt{3}}{2}k_{2})].
\end{eqnarray}
Making $k_{2}$ complex leads to interactions
on a layered triangular lattice etc.

\section{Higher  $O(n>1)$ models And Symmetries Intermediate
Between Global and Local}
\label{high}

The multi-component $O(n>1)$ spin
models display a far richer 
variety of possible higher
dimensional 
extensions.

The constant $\lambda$ need not be only a
complex number, it may also be 
quaternion or correspond 
to more general set of matrices.

Let us examine the extension 
to ``quaternion'' $\vec{k}$
(or with the matrices
$i\sigma_{1},i\sigma_{2}$,and $i\sigma_{3}$
(with $\{ \sigma_{\alpha} \}$ the Pauli matrices) 
taking on the role of $i,j$ and $k$).
If we were dealing with an $O(2)$ (or XY) model,
the same high temperature expansion 
could be reproduced:

We may envisage a trivial extension to the usual 
scalar product Hamiltonian
\begin{equation}
H= \frac{1}{2}\sum_{\vec{x},\vec{y}} \hat{V}(\vec{x}-\vec{y})
\vec{S}(\vec{x}) \cdot \vec{S}(\vec{y})  
\end{equation}
to one in which the kernel is no
longer diagonal in the internal 
spin indices $\alpha,\beta  = 1,2$.

\begin{equation}
H= \frac{1}{2}\sum_{\vec{x},\vec{y}} \hat{V}_{\alpha,\beta}(\vec{x}-\vec{y})
S_{\alpha}(\vec{x})  S_{\beta}(\vec{y}).  
\end{equation}

The Hubbard Stratonovich transformation
for $Z = Tr \{ \exp[-\beta H] \}$ 
with the bilinear $ H = \frac{1}{2} \sum_{\kappa \rho} S_{\kappa}
\hat{V}_{\kappa \rho} S_{\rho}$
proceeds as before. The super-index $\kappa = (\vec{x}, \alpha)$
now labels both the physical coordinate $\vec{x}$ and the 
polarization $\alpha$. And, just as
before, the trace of 
$\exp[ i S_{\kappa} \eta_{\kappa}]$ over all $\{S_{\kappa}\}$ is
$\prod_{\vec{x}} I_{n/2-1}(|\eta(\vec{x}|)$.
The high temperature expansion 
of section (\ref{a}) may be reproduced
word for word with the  
kernel $V_{\alpha,\beta}(\vec{x}-\vec{y})$ 
now generated by each contraction of 
$\eta_{\alpha}(\vec{x})$ and $\eta_{\beta}(\vec{y})$,
or in momentum space $v_{\alpha,\beta}(\vec{k}) 
\delta_{\vec{k}+\vec{k}^{\prime},0}$ 
is generated  by the contraction of 
$\eta_{\alpha}(\vec{k})$ and 
$\eta_{\beta}(\vec{k}^{\prime})$.
In the evaluation of the 
free energy or the partition function, a 
given bubble diagram reads
\begin{eqnarray}
 \Big( \int \prod_{b=1}^{\mbox{loops}} \frac{d^{d}k_{b}}{(2 \pi)^{d}}
\sum_{l=1}^{d}  Tr \{ \exp[i \sum_{a,b} M_{ab} k_{b}^{l}] \} \Big)
\end{eqnarray}
where, as before, $a$ spans the number of propagators,
$b$ spans the number of loops, and $k^{l}_{b}$ with 
$1 \le l \le d$ denotes the $l$-th component
of the $d-$dimensional wave-vector within the b-th
loop $\vec{k}_{b}$.

\bigskip

Let us start off with a nearest neighbor one dimensional 
XY chain and consider the transformation
\begin{eqnarray}
\exp[i k_{1}] \rightarrow \exp[i (k_{1} + i 
\lambda_{1} k_{2} \sigma_{1} + i \lambda_{2} k_{3} \sigma_{2}
+ i \lambda_{3} k_{4} \sigma_{3})] \nonumber
\\ \equiv \exp[i k_{1} - \vec{w} 
\cdot \vec{\sigma}].
\end{eqnarray}
In the argument of the exponential the identity matrix 
$1$ commutes with $\vec{w} \cdot
\vec{\sigma}$ and the 
exponential may be trivially
expanded as 
\begin{eqnarray}
\exp[i k_{1} - \vec{w} \cdot
\vec{\sigma}] = \exp[i k_{1}] \exp[-  \vec{w} \cdot
\vec{\sigma}].
\end{eqnarray}
Once again in higher dimensions 
(this time $d=4$) the $k_{1}$ integration will
reproduce the familiar 
\begin{equation}
\int_{-\pi}^{\pi} dk_{1} \exp[i N^{\prime} k_{1}] = 2 \pi \delta_{N^{\prime},0}
\label{fam}
\end{equation}
for each independent loop momentum $\vec{k}$
and the integration over the remaining 
$k_{2},k_{3},$ and $k_{4}$ components
will simply yield multiplicative 
constants.

Thus the kernel  
\begin{equation}
v(\vec{k}) = \exp[i k_{1} - \vec{w} \cdot
\vec{\sigma}] + h.c. 
\end{equation}
will generate a four dimensional
XY model which has the partition 
function of a nearest neighbor one dimensional 
XY chain. The corresponding 
real space kernel
\begin{eqnarray}
\hat{V}_{\alpha \beta}
(x_{1},x_{2},x_{3},x_{4}) = \int_{-\pi}^{\pi} ... \int_{-\pi}^{\pi} 
\frac{d^{4}k}{(2 \pi)^{4}} \exp[i \vec{k} \cdot \vec{x}]  
\Big(\exp[i (k_{1}  \nonumber
\\ + i \lambda_{1} k_{2} \sigma_{1} + 
i \lambda_{2} k_{3} \sigma_{2} 
 + i \lambda_{3} k_{4} \sigma_{3})]_{\alpha \beta} + h.c. \Big).
\nonumber
\end{eqnarray}
In general, this kernel is no longer 
diagonal in the internal 
spin coordinates (unless $\vec{w}$ happens
to be oriented along $\sigma_{3}$).
\begin{eqnarray}
|\vec{w}| = \sum_{i=2}^{4} \lambda_{i-1}^{2} k_{i}^{2} \nonumber
\\ \hat{w} = (\frac{\lambda_{1} k_{2}}{|\vec{w}|},  \frac{\lambda_{2}
k_{3}}{|\vec{w}|}, \frac{\lambda_{3} k_{4}}{|\vec{w}|})
\end{eqnarray}
and 
\begin{eqnarray}
\exp[-\vec{\sigma} \cdot \vec{w}] = \cosh (|\vec{w}|) - \sinh
(|\vec{w}|) (\vec{\sigma} \cdot \hat{w})
\end{eqnarray}
For general $\{ \lambda_{i} \}_{i=1}^{3}$ 
the inverse Fourier transform is nontrivial.
The scalar product is replaced
by a more complicated product
amongst the components. 
The real space interaction
kernel is once again
algebraically long ranged
along the $x_{2},x_{2}$ and $x_{4}$
axes.

Just as quaternions
(or Pauli matrices $\{ \sigma_{i} \}$)
may be employed
we may also consider the 
transformation
\begin{eqnarray}
k_{1} \to k_{1} + k_{\mu+2} \gamma^{\mu}
\end{eqnarray}
with the Dirac matrices 
$\{ \gamma^{\mu} \}$,
with its associated
propagator 
$\exp[i (k_{1}+ k_{\mu+2} \gamma^{\mu})]$
to produce solvable models in $2 \le d \le 5$
dimensions (depending on how many 
momentum components are contracted
with the gamma matrices).
The partition function of these models
is identically that of a nearest
neighbor one dimensional system.
An infinite number of variations
along this theme are possible.

\bigskip

\bigskip

In this manner we may generate
a multitude of solvable, slightly 
more realistic, models with a 
{\bf real hermitian} kernel.

The hermitian kernel
\begin{eqnarray}
\hat{v}(\vec{k}) = -2 \cos(k_{1} +  k_{2} \sigma_{3})
\label{O_2}
\end{eqnarray}
leads to the partition function of 
a nearest neighbor one dimensional 
$O(2)$ chain.

\bigskip

The proof amounts to a reproduction of the
calculations given above for the four 
dimensional $O(2)$ system.

\bigskip

In real space the latter kernel (Eqn.(\ref{O_2}))
leads to the $O(2)$ 
Hamiltonian 
\begin{eqnarray}
H = - \sum_{\langle i j \rangle~ along~(1,1)} S_{i}^{(1)} S_{j}^{(1)} 
\nonumber
\\
 - \sum_{\langle i j \rangle~ along~(1,-1)} S_{i}^{(2)} S_{j}^{(2)} 
\label{dec}
\end{eqnarray}
where the $n=1$ (or x component) 
of the spins interact in the first 
term and only the y-components
of the XY spins appear in the 
second term; the indices $i$ and
$j$ are the two dimensional 
square lattice coordinates.
The first term corresponds to
interactions along the $(1,1)$
direction in the plane
and the second term corresponds
to interactions along the 
$(1,-1)$ diagonal. 
The two spin components
satisfy $[S_{i}^{(1)}]^{2}+ [S_{i}^{(2)}]^{2}=1$-
i.e. are normalized at every site $i$.
By our mapping, this model 
trivially has the partition function
of a one dimensional nearest neighbor $O(2)$ spin 
chain. Thus its free energy 
per site $f = - \beta^{-1} \ln [I_{0}(\beta)]$
where $I_{0}$ denotes the associated 
Bessel function.  

It amusing to trivially note 
by inspection that this explicitly works for works the
periodic 2 $\times$ 2 system on the torus where the two diagonals
decouple and along each diagonal one has the interactions 
of the standard $O(2)$ symmetric one dimensional XY model.

This system is frustrated-  glancing at the Hamiltonian of
Eqn.(\ref{dec}) we note, that it 
is impossible to saturate the bonds associated
with both the $x$ and $y$ polarizations 
of the spins simultaneously. The state in 
which the $x$ component interactions
are minimized  is geometrically
orthogonal to one in which the $y-$component
interactions are minimized.
In fact, we just saw that the frustration is so large 
that it thwarts the generic low temperature two dimensional quasi (algebraic) 
long range order and leads to a canonical one dimensional like 
behavior instead. 

We may also generate interactions
along arbitrary tilted rays
by considering 
\begin{eqnarray}
\hat{v}(\vec{k}) = -2 \cos(k_{1} +  \lambda k_{2} \sigma_{3})
\label{O_2}
\end{eqnarray}
to show that the 
Hamiltonian
\begin{eqnarray}
H = - \sum_{\langle i j \rangle~ along~(1,\lambda)}  S_{i}^{(1)}
S_{j}^{(1)} \nonumber
 \\
- \sum_{\langle i j \rangle~ along~(1,-\lambda)}  S_{i}^{(2)} S_{j}^{(2)}\label{super-d++}
\end{eqnarray}
is equivalent to 
\begin{eqnarray}
H = - \sum_{\langle i j \rangle}  \vec{S}_{i} \cdot
\vec{S}_{j}.  
\label{super-c++}
\end{eqnarray}
By tuning $\lambda \to 0$, the interactions
along the two ``clapping'' 
directions $(1, \pm \lambda)$ fold back and degenerate
into the original one dimensional, 
$O(2)$ symmetric, chain.

Similarly, by choosing the diagonal 
\begin{eqnarray}
\hat{v}(\vec{k}) = - 2 \cos(k_{1} P_{+}+ k_{2} P_{-}),
\end{eqnarray}
with the projection operators
$P_{\pm} \equiv \frac{1}{2} (1 \pm \sigma_{3})$,
we may prove that 
\begin{eqnarray}
H = - \sum_{\langle i j \rangle~ along~(1,0)} \ S_{i}^{(1)}
S_{j}^{(1)} \nonumber
 \\
- \sum_{\langle i j \rangle~ along~(0,1)}  S_{i}^{(2)}
S_{j}^{(2)}\label{super-e++}
\end{eqnarray} 
is equivalent to the one dimensional
\begin{eqnarray}
H = - \sum_{\langle i j \rangle}  \vec{S}_{i} \cdot
\vec{S}_{j}.  
\label{super-f++}
\end{eqnarray}
Here as all $\hat{v}$ matrices
are diagonal in the internal 
spin indices, the 
traces of the two dimensional 
system (Eqn.(\ref{super-e++})) are 
identically the same as those 
of the one dimensional system
of Eqn.(\ref{super-f++}).
Schematically, in the two dimensional system,
for each given loop momenta $\vec{k}^{b}$ having the 
two spatial
components ($k_{1}^{b}, k_{2}^{b}$), the integrals
are exactly the same as they are
for the one dimensional case: 
\begin{displaymath}
Tr \int \frac{dk_{1}^{b}}{2 \pi} \int \frac{dk_{2}^{b}}{2 \pi}
\left[ \left( \begin{array}{ccc}
e^{i
k_{1}^{b}N^{\prime}} ... & ~ 0 \\
0 &  e^{i
k_{2}^{b}N^{\prime}} ...
\end{array} \right) ... \right]  
\end{displaymath}
\begin{displaymath}
=  Tr \int \frac{dk_{1}^{b}}{2 \pi} 
\left[ \left( \begin{array}{ccc}
 e^{i
k_{1}^{b}N^{\prime}} ... & ~ 0 \\
0 &  e^{i
k_{1}^{b}N^{\prime}} ...
\end{array} \right)... \right].
\end{displaymath}

After integrations, each one of the 
identical diagonal entries will be given 
by Eqs.(\ref{color}). These are the exactly 
the same integrals that we have evaluated for 
the Ising case. (As told, it is merely
the weights of the 
vertices which change
in the general $O(n)$ 
model.)  The dimensional 
reduction of Eqn.(\ref{super-e++}) 
is ambivalent to the special 
nearest neighbor short range (or even 
lattice character) of the problem. If the 
x-components of the spins interact
with a kernel which is $x_{1}$
direction dependent and the y-components
would interact with the same kernel
along the $x_{2}$ direction then 
a dimensional reduction would be possible.
In general, and in the continuum
in particular, one may, of course, perform
rotations within the external
momentum space and internal
spin space in order to 
examine dimensional 
reductions for all 
interaction directions.

As told, on the lattice, all of this need not apply to special short
range interactions.
More generally, we may replace $k_{1} \to k_{1} + k_{2} \sigma_{3}$
within an arbitrary kernel
\begin{eqnarray}
\hat{v}(k_{1}) = - 2 \sum_{r=1}^{R} v_{r} \cos(r k_{1})
\end{eqnarray}
of spatial range $R \ge 1$. Employing Eqn.(\ref{fam})
once again we note that any XY spin system with a Hamiltonian
\begin{eqnarray}
H = - \sum_{\langle i j \rangle~ along~(1,1)} \hat{V}_{ij} S_{i}^{(1)}
S_{j}^{(1)} \nonumber
 \\
- \sum_{\langle i j \rangle~ along~(1,-1)} \hat{V}_{ij} S_{i}^{(2)} S_{j}^{(2)}\label{super-dec}
\end{eqnarray}
with a general translationally
invariant $\hat{V}_{ij} = f(|\vec{i} - \vec{j}|)$ 
of an arbitrary range $R$ (i.e. one which does not 
necessarily link only nearest 
neighbor diagonal sites $i$ and $j$) 
has exactly the same
partition function 
of a one dimensional
system whose Hamiltonian
\begin{eqnarray}
H = - \sum_{\langle i j \rangle} \hat{V}_{ij}~ \vec{S}_{i} \cdot
\vec{S}_{j}
\label{super-comp}
\end{eqnarray}
with the arbitrary range interaction
kernel $\hat{V}_{ij}$.

\bigskip

Similarly, the partition
function of an XY system
having the three dimensional
kernel $\hat{v}(\vec{k}) = -2 \cos k_{1} - 2 \cos (k_{2}+k_{3} \sigma_{3})$
is equal to that of the two dimensional nearest neighbor XY ferromagnet
which exhibits Kosterlitz-Thouless like behavior at sufficiently
large inverse temperature $\beta$ (or temperatures $T< T_{KT}$).
Here each of two spin components interacts within a different
subplane. The coupling between the two spin 
components due to the normalization
constraint apparently plays no role
in leaving the system two dimensional.

\bigskip

In a similar manner,  any $d^{\prime}$
dimensional spin system (at temperatures  $ T> T_{c}$) 
can be made have an effective
dimensionality $1 \le d \le d^{\prime}$
without the physical need of
actual geometrical compactification.

\bigskip

The models above display special 
symmetries, intermediate between those
of the standard two spin $O(n)$ models which 
display a global group symmetry
and lattice gauge theories
which a local group symmetry at 
every lattice site. 
Note that the transformation 
\begin{eqnarray}
\left\{  \begin{array}{ll} 
S_{\alpha=1}^{(x_{1},x_{2})}  \to \eta_{x_{1}}
S_{\alpha=1}^{(x_{1},x_{2})} \\
S_{\alpha=2}^{(x_{1},x_{2})}  \to \eta_{x_{2}}
S_{\alpha=2}^{(x_{1},x_{2})}
\label{spec-symm}
\end{array} \right.
\end{eqnarray}
with the $(2L)$ arbitrary, ``gauge'', degrees of 
freedom $\eta_{x_{i}} = \pm 1$
leaves Eqn.(\ref{super-e++}) invariant.
Here $(x_{1},x_{2})$ denote the two spatial components
of the spin coordinate $\vec{x}$. Note that by this
symmetry, any particular state (including the ground state) 
is, at least, $2^{2L}$ degenerate- yet another manifestation
of ``frustration''.
The effective gauge redundancy 
of Eqn.(\ref{spec-symm})
is midway between a full
blown gauge theory (where 
redundant gauge degrees
of freedom live on every site)
and a globally symmetric theory 
where only one global symmetry exists.
The number of gauge degrees of freedom
in Eqn.(\ref{spec-symm}) is the same 
as that in a one dimensional
$Z_{2}$ gauge system. In general, 
the manifold of the gauge degrees of freedom
can assume any effective dimensionality
in between the two standard extremes- $d$ (gauge theories) 
and zero (globally symmetric spin models). 
The 
$O(4)$ spin model 
\begin{eqnarray}
H = - \sum_{\langle i j \rangle \mbox{ along (1,1)}}  (S_{i}^{(1)}
S_{j}^{(1)} + S_{i}^{(3)}
S_{j}^{(3)})  \nonumber
\\ - \sum_{\langle i j \rangle \mbox{ along (1,-1)}}  (S_{i}^{(2)}
S_{j}^{(2)} + S_{i}^{(4)}
S_{j}^{(4)})  
\end{eqnarray}
displays $(2L)$ independent $SO(2)$ (or $U(1)$) gauge degrees of 
freedom $\eta$ residing on
each individual diagonal. Explicitly
\begin{eqnarray}
\left\{  \begin{array}{ll} 
S_{\alpha=1}^{(x_{1},x_{2})} + iS_{\alpha=3}^{(x_{1},x_{2})}    
\to \eta_{d_{1}} (S_{\alpha=1}^{(x_{1},x_{2})} +
iS_{\alpha=3}^{(x_{1},x_{2})}) \\
S_{\alpha=2}^{(x_{1},x_{2})}  + iS_{\alpha=4}^{(x_{1},x_{2})}  \to \eta_{d_{2}}
(S_{\alpha=2}^{(x_{1},x_{2})} + iS_{\alpha=4}^{(x_{1},x_{2})}),
\label{spec-symm+}
\end{array} \right.
\end{eqnarray}
with arbitrary phases $\eta_{d_{i}}$ along the $2L$ diagonals $d_{1,2}$, 
is a symmetry. By our mapping, this two dimensional system 
with $\sim L$ redundant $O(2)$ gauge degrees of freedom is equivalent to an isotropic,
nearest neighbor, $O(4)$ spin chain having
only a single zero dimensional global symmetry. 
The variations are endless.

Although we have focused attention
on equivalence to simple models 
with rotationally invariant scalar 
products (e .g. Eqn.(\ref{super-comp})), much 
of what we said is also generally
valid for non-isotropic 
actions \cite{z-z}. 

Finally. in order to set the stage
for the discussion of a 
cartoon of directed electronic orbitals (e.g. $p$ or $d$ 
orbitals appearing in some systems (e.g. the manganates),
we will now introduce ``directed clock
models''. These models are just the 
standard $\hat{N}-$state clock models
with the additional twist
that an interaction (``hopping'')
can occur only along an axis 
parallel to the spin direction
(a cartoon for the direction along which 
the electronic orbital is extended).
The high temperature diagrammatic 
expansion for these models may proceed
as before, with the provision
that the generating function
(which merely sets the weights of
all vertices) is no longer
a rotationally invariant Bessel
function but rather $\sum_{i=1}^{\tilde{N}} \exp[\vec{S}_{i} \cdot 
\vec{\eta}]$. For the directed 4-state 
clock model, the four possible 
spin orientations
are $(\pm 1,0)$ and
$(0, \pm 1)$ and the 
Hamiltonian is none other 
than that of Eqn.(\ref{super-e++}). 
Our mapping will show 
that this is equivalent to
the standard (undirected)
one dimensional nearest neighbor 4-state
clock model whose 
partition function is 
trivially twice of 
that the nearest neighbor 
one dimensional Ising 
model. The generating
function is 
\begin{eqnarray}
\langle \prod_{\vec{x}} [2(\cosh \eta_{1} + \cosh
\eta_{2})] \rangle \to \nonumber
\\ \langle \prod_{\vec{x}} [2
\cosh(\eta_{+})] \rangle   \langle \prod_{\vec{x}} [2
\cosh(\eta_{-})] \rangle,  
\end{eqnarray}
with $\eta_{\pm} \equiv (\eta_{1} \pm \eta_{2})/2$.
The partition function is a direct product 
of the partition function terms for 
two independent Ising systems.
This has a much clearer intuitive
meaning (see Appendix B)- the direct product 
space of two Ising spins sitting
at every site gives rise to
two dimensional spins
whose components are
$(\pm 1, \pm 1)$. Upon rotation
and normalization, this becomes
the four state clock 
model just discussed.

In general, it is easy to see
that the partition function
of the $(2^{d})$ state directed 
clock model (in which each spin
has $d$ components) embedded in 
an external space of $d$ dimensions
is equivalent to that of the
one dimensional Ising
model (more precisely,
$d$ replicas of Ising
systems).

We may similarly address \cite{z-z} the 
``squared directed 4-state
clock model''
\begin{eqnarray}
H = - \sum_{\langle i j \rangle~ along~(1,0)}  [S_{i}^{(1)}
S_{j}^{(1)}]^{2} \nonumber
 \\
- \sum_{\langle i j \rangle~ along~(0,1)}  [S_{i}^{(2)}
S_{j}^{(2)}]^{2}
\end{eqnarray}
to show that its partition
function is equivalent to
that of a one-dimensional
model with 
\begin{eqnarray}
H = - \sum_{i} S_{i}^{2} S_{i+1}^{2} = const~ (=-N),
\end{eqnarray}
which is trivial.

The reader is referred to Appendix B 
for a perspective on the conventional, non-directed,
and directed clock models in $n \ge 2$  
internal spin 
dimensions.

\section{Potential Physical Realizations}

Some frustrated systems might have
realizations close to the 
special symmetry points
at which we proved the 
occurence of exact dimensional
reductions.

\subsection{The Manganates}

\begin{figure}
\begin{center}
 \epsfig{figure=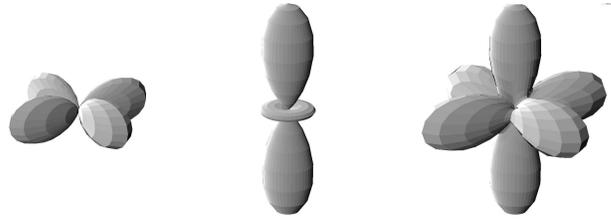,width=0.95\linewidth}
\end{center}
\caption[]{Left and middle: the conventional real space
functions $\phi_{a} = x^{2}-y^{2}$, $\phi_{b} = 3z^{2}-r^{2}$
for the cubic E doublet. Right: the complex 
combinations $\phi_{a} \pm i \phi_{b}$ are
octupolar states. \cite{Fazekas}}
\label{cub}
\end{figure}

Before giving a superficial
flavor of one potential
system where models
and symmetries such
as those introduced
in the previous section
might be of real practical
relevance, let us quickly 
outline the logic. 
As seen in the central figure of Fig.(\ref{cub}),
orbital wave functions
can be cigar like 
and highly direction. In such
a case, overlap integrals 
will be much larger for 
hopping or interactions
along one axis than
along another. Hopping
and/or spin-orbit couplings 
might depend on the 
state of the orbital
direction (``polarization'').

Polarization dependent interactions 
akin to those in Eqn.(\ref{dec}) may be of relevance in discussing 
directed orbital problems 
in two dimensions (e.g. those associated 
with the electronic orbitals in 
transition metal oxides of perovskite 
structure) \cite{tokura}.
The non-isotropic
electronic orbitals lead
to overlap terms which 
are strongly direction 
dependent. In these materials
(e.g. La$_{2-x}$Sr$_{x}$MnO$_{3}$),
the surrounding oxygen atoms introduce crystal
field splitting of the  $d$ 
orbitals of the caged 
transition metal (e.g. Mn).
Wave functions pointing towards 
the oxygen ions may attain higher 
energy than those that point 
in between. The two ``bad'' higher energy $d$ orbitals
are $|x^{2}-y^{2} \rangle$ and $|3z^{2}-r^{2}\rangle$
(customarily denoted as e$_{g}$ orbitals) are depicted in
Fig.(\ref{cub}).
The ground state is determined by Hund's rule. A 
Mn$^{3+}$ ion, for example, has a d$^{4}$ 
configuration and consequently, one electron 
populates on the states shown in Fig.(\ref{cub}).
These states dominate the orbital physics.

The directed four or six state 
clock models may be 
regarded as oversimplified
cartoons for the orbital lobes
in systems such as these. 
The overlap terms allow 
hopping only in the direction
lying along the axis of symmetry
of the orbital.
They are a very crude
model for 
\begin{eqnarray}
H = \sum_{\langle ij \rangle} 
t_{ij} (c_{i \sigma}^{\dagger} c_{j \sigma} + h.c.)
\end{eqnarray}
where the hopping amplitudes $\{t_{ij}\}$ 
depend on internal indices indicating
the relative orientation between the 
orbitals at sites $i$ and $j$.
The hopping amplitudes between two neighboring
Mn atoms depend on the overlap of the 
d-orbitals with the oxygen p-orbitals
which lie in between.

Another origin of such
a behavior may result
from spin-orbit
couplings.

\subsection{Frustrations and Transitions within
the checkerboard and Pyrochlore lattices}

As demonstrated by E. Berg et al. \cite{erez},
the S = 1/2 Heisenberg antiferromagnet on the 
frustrated checkerboard and pyrochlore 
lattices, can be related to
models of the form similar 
to those that we have looked at
till now. 

\begin{figure}
\begin{center}
 \epsfig{figure=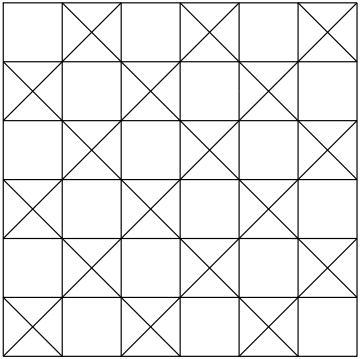,width=0.55\linewidth}
\end{center}
\caption[]{The checkerboard lattice.
The two singlet ground states of the 
crossed plaquettes may be represented
by an Ising pseudospin lying along
the horizontal axis. The effective
interaction amogst these pseudo-spins 
is encapsulated by Eqn.(\ref{EREZ}).}
\label{checker}
\end{figure}

For instance \cite{erez}, in the low energy regime, 
the Heisenberg model on the
checkerboard lattice shown in Fig.(\ref{checker}), 
is related to the Ising like model 
\begin{eqnarray}
H = - J \sum_{\langle i j \rangle} (\vec{S}_{i} \cdot \hat{e}_{ij}) 
(\vec{S}_{j} \cdot \hat{e}_{ij}) - h\sum_{i} S_{i}^{x}.
\label{EREZ}
\end{eqnarray}
Here the directors $\hat{e}_{ij}$ may
point along directions 
at angles $\pi/3 (-\pi/3)$ away from
the x axis for horizontal (veritcal)
bonds respectively. The spins correspond to
vertical or horizontal dimers\cite{erez}.
Had the directors $\hat{e}_{ij}$ been along 
orthogonal axis, we would just obtain a trivial
counterpart of the Hamiltonian of 
Eqn.(\ref{dec}). At the orthogonal 
point (if the directors 
were orthogonal), the system is exactly
be one dimensional. Sans the 
applied mangetic field, the free energy density
at this point is $-\beta^{-1} \ln \cosh \beta$ times
a combinatorical factor equal to the 
logartihm of the number of ways 
in which the plane may be tiled with
horizontal and vertical lines 
such that each point belongs to
exactly one line. 

A deviation of $\{\hat{e}_{ij}\}$ from 
orthogonality, leads to 
the removal of the degneracy
associated with the symmetries
of Eqn.(\ref{spec-symm}) and
its likes, and the system is 
indeed expected to display 
canonical two dimensional behavior
and have a transition 
at a finite temperature.

A similar model 
may be written for the
pyrochlore lattice \cite{erez}.

Thus, some geomertically
frusrated models are
close to yet removed
from the special symmetry
points where dimensional
reduction would occur.

\section{Special Two Dimensional Bethe Ansatz solutions}
\label{Bethe}

We may consider quantum (spin, fermionic, bosonic) 
extensions in which the unperturbed action $S_{0}$ 
links two nearest neighbor sites. By the Trotter
formula, the partition function may be 
written as a lattice sum over a $(d+1)$ 
dimensional system of thickness $\beta$
(the inverse temperature) along the imaginary time axis (the $(d+1)$-th 
direction). The $(d+1)$-th momentum component $k_{d+1}$
is Matsubara frequency $\omega$ whose allowed values
$\{ \omega_{n} \}$ are dictated by the periodic/antiperiodic 
boundary conditions along the imaginary 
time axis that must be imposed for a bosonic/fermionic
system respectively. The Green's function conjugate to 
$S_{0}$ in the undualized (non Hubbard-Stratonovich
transformed) system reads
\begin{eqnarray}
G_{0} = \Big[\beta \Big( -2 \sum_{l=1}^{d+1} \cos k_{l} + A \Big)\Big]^{-1}.
\end{eqnarray}

If the magnitude of the squared mass $|m_{0}|^{2} =|A| =const > 2(d+1)$
then a geometric series expansion 
in A$^{-1}$ may be performed
to generate a sum of harmonics.
Within each diagrammatic term,
we may take $k_{1} \to k_{1} + i \lambda k_{2}$ 
or other similar extensions to 
create high dimensional
problems that have
the same partition 
function as that 
of a solvable 
canonical one 
dimensional
system.

For a relativistic spin-1/2 system, we may employ the naive 
fermionic propagator
\begin{eqnarray}
G_{0} = [\beta(\gamma_{\mu} \sin k^{\mu} + m_{0})]^{-1}
\end{eqnarray}
and once again expand the resulting 
diagram in powers of $m_{0}^{-1}$
to obtain a sum of harmonics. 
This holds for all canonical lattice 
theories in which the kernel are periodic
functions of the momenta and thus, when well behaved,
may be expressed as sums
of harmonics.

Eqn.(\ref{fam}) forces all
diagrams with substitutions 
such as $k_{1} \to k_{1} + i \lambda k_{2}$ 
to attain the same value 
as they would attain in 
the original system.

If an exact Bethe Ansatz solution is known
for a fermionic one dimensional system
with nearest neighbor hopping 
then in some cases it may
be extended to two dimensions where 
the hopping matrix element $K(x_{1},x_{2})$
for separation $\vec{x}=(x_{1},x_{2})$
attains exactly the same form as the two-dimensional
kernel $\hat{V}(x_{1},x_{2})$ just given previously.
Loosely speaking, if in a fictitious electronic system
the hopping matrix element would be the non hermitian 
\begin{eqnarray}
K(x_{1},x_{2}) = 
 - \frac{1}{\pi} \Big( (-1)^{x_{2}} \frac{\sinh \pi \lambda_{1}}{\lambda_{1}^{2}+ x_{2}^{2}}
 \delta_{x_{1},0} \nonumber
\\ 
\Big[\lambda_{1}  (\delta_{x_{2},1}+ \delta_{x_{2},-1}) \nonumber
\\ + i x_{3} ( \delta_{x_{2},1}- \delta_{x_{2},-1})
 \Big] \Big)
\end{eqnarray}
then the system would essentially be
a one dimensional Luttinger liquid
along $x_{1}$. If the kernel would instead decay 
algebraically along the imaginary 
time axis the problem would become 
that of dissipating system.

A hermitian hopping matrix 
element could also do the trick if we 
were to consider polarization
dependent hopping (or ``interactions'') 
analogous to those in Eqn.(\ref{dec})
and slightly more complicated
variants. As noted earlier, 
such models and the more general
equivalence of the two dimensional
system of Eqn.(\ref{super-dec}) with the one dimensional
one of Eqn.(\ref{super-comp}) may be of relevance 
in discussing directed orbital problems in two 
dimensions in systems like 
the manganates \cite{tokura}.

\section{Dimensional Reduction In Lattice Gauge Theories}
\label{GAUGE}

Similarly, the standard lattice gauge theories
may be mapped to higher dimensional lattice
gauge theories. 
Let us further show also the opposite-
the dimensional
reduction that trivially
occurs in going
from a two dimensional
gauge theory to a one
dimensional spin chain
is one special
case in our general
way of constructing 
models of different
dimensions that are 
equivalent.

A pure lattice gauge theories
(for any group) may be mapped
onto a Coulomb gas model with 
nontrivial onsite interactions.
The advantage is that 
in the Coulomb gas all
interactions are between
pairs and our
machinery for dimensional
reduction
can be brought to 
bear. To avoid carrying too many
indices around
we will now consider the
Abelian $Z_{2}$ (Ising) 
group.

Let us consider the even sublattice
sites 
\begin{eqnarray}
\sum_{a=1}^{d} x_{a} \equiv 0 (\mbox{mod 2})
\end{eqnarray}
in a hypercubic lattice, where $x_{a}$ labels the 
$a$-th component of the position vector of a site
on the lattice. On this sublattice dual
fields $\zeta_{i}$ are introduced. 

Each bond variable $U_{ij}$ is connected
to one and exactly one lattice site ($i$ or $j$) which 
belongs to the even sublattice (where the fields $\zeta$
exist). The partition function may then 
be schematically written as 

\begin{eqnarray}
 Z = \int DU \exp[K \sum_{\Box} UUUU] \nonumber
\\ \equiv \int DU \prod_{\alpha \beta \gamma \delta} 
\exp[\frac{1}{2} V_{\alpha
\beta;\gamma \delta} U_{\alpha} U_{\beta} U_{\gamma} U_{\delta}] \nonumber
\\ = {\cal{N}} \int DU \int D\zeta \exp[\frac{1}{2} V^{-1}_{\alpha
\beta \gamma \delta} \zeta_{\alpha \beta} \zeta_{\gamma \delta}  -
i \zeta_{\alpha \beta} U_{\alpha} U_{\beta}]
\end{eqnarray}
where the kernel $V_{\alpha \beta; \gamma \delta} = K \delta_{|\alpha
\beta- \gamma \delta|,\sqrt{2}}$ where the Greek indices denote
the bonds and $\alpha \beta$ denotes the site common
to bonds $\alpha$ and $\beta$. The distance between
two sites was marked by $|\alpha
\beta- \gamma \delta|$. At each sublattice site
$i$ there are $2d(d-1)$ fields
$\{\zeta_{\alpha \beta}\}$. As each link $U_{ij}$
has only a single end point $i$ which lies
in the even sublattice, only a dependence on one
point $\vec{\zeta}_{i}$ will remain when integrating (or tracing) 
over the link variables that have one end point
at site $i$,
\begin{eqnarray}
\int D U \exp[-i U \zeta U] = \prod_{i} f(\vec{\zeta}_{i}) 
\label{on-f}
\end{eqnarray}
where the $2d(d-1)$ fields have been lumped
into the vector $\vec{\zeta}_{i}$. 
The effective dual action 
\begin{eqnarray}
S_{eff} = - \frac{1}{2} V_{\alpha \beta; \gamma \delta}^{-1} \zeta_{\alpha
\beta} \zeta_{\gamma \delta} - \sum_{i} \ln f(\vec{\zeta}_{i}).
\end{eqnarray}
This is a $2d(d-1)$ component spin system with 
pair interactions and a nontrivial onsite interaction. 
As $V$ corresponds to nearest neighbor interaction
on the even sublattice ($\sim k^{2}$ for small
wave numbers), 
its inverse corresponds to Coulombic interactions ($\sim k^{-2}$)
in the long wavelength limit. More precisely, the dual model 
corresponds to coupled Coulomb gases
having a net of 2d(d+1) components
(in the Abelian case) instead of
one. 

The function $f(\vec{\zeta}_{i})$ is always even and is further
always invariant under the group symmetry of $U$. In the argument
of the exponent of Eqn.(\ref{on-f}) 
\begin{eqnarray}
Tr\Big[ (gUg^{-1} gUg^{-1}\zeta)^{p} \Big] =
Tr\Big[(gUUg^{-1}\zeta)^{p}\Big] \nonumber
\\ = Tr\Big[(g^{-1}\zeta g UU)^{p} \Big]
\end{eqnarray}
for any integer power $p \ge 1$ in the 
expansion, where the cyclic property of the trace and the 
group invariance of the measure $DU$ were employed. 
A pure gauge theory is equivalent
to gauge invariant charges interacting
by pair Coulomb potentials.

As the two dimensional $Z_{2}$ lattice 
gauge theory is equivalent to
the one dimensional Ising chain,
its dual, a special isotropic two dimensional 
Coulomb gas with spin-spin  
interactions, has 
a canonical one dimensional
behavior. (Its partition
function $Z = [(\cosh K)^{L} +  (\sinh K)^{L}]^{L}$
(with $L$ the linear extent of the system) if the two dimensional system
has periodic boundary conditions along both axis (each of length $L$)). 
This is a simple example of a dimensional reduction. The same applies to a 
$U(1)$ system where in the two dimensional dual model-
a special, eight component XY type model with
Coulomb like interactions-
we will find that the free energy
density $- N^{-1} \ln Z$ tends to 
$(-\ln[I_{0}(K)])$ with $I_{0}$
a Bessel function.

In fact, by selecting the temporal
gauge, in which the gauge links
$U_{0}(n)=1$ along all links along the time (0) direction,
any two (or 1+1) dimensional gauge theory with plaquette terms 
can be recast as a classical one dimensional model
having nearest neighbor interactions.
If we dualize any two dimensional gauge theory
we obtain a special, spatially symmetric, two 
dimensional Coulomb gas that is a nearest neighbor 
one dimensional system in disguise.

To get an appreciation for
what is happening, let 
us examine matters in
detail. For the
$Z_{2}$ lattice gauge
theory in $d=2$ dimensions,
the vector $\zeta$ is four-dimensional
and at each site, 
\begin{eqnarray}
f(\vec{\zeta}) = 16( \cos \zeta_{1} \cos \zeta_{2} \cos \zeta_{3}
\cos \zeta_{4} \nonumber
\\ + \sin \zeta_{1} \sin \zeta_{2} \sin \zeta_{3}  
\sin \zeta_{4}),
\end{eqnarray} 
where we have marked in a counterclockwise
fashion the bonds $\{ U \}$ about a given site $i$
by $U_{1}, U_{2}, U_{3}$ and $U_{4}$.
With these bonds we identify 4 dual fields
in the following way- $\zeta_{1}$ arises
from a coupling to the two bond product $U_{4} U_{1}$,
$\zeta_{2}$ arises
from a coupling to the product $U_{1} U_{2}$,
and so on cyclically.  

The interaction 
kernel in the 4 dimensional internal
space of $\vec{\zeta}(\vec{k})$ 
(the components of $\zeta$ in 
internal and momentum space), reads
\begin{eqnarray}
v = - K \left( \begin{array}{cc}
0 & A_{2} \\
A_{2} & 0
\end{array} \right), 
\end{eqnarray}
with the two by two submatrix
\begin{eqnarray}
A_{2} = \left( \begin{array}{cc}
\cos (k_{1}+ k_{2})  & 0\\
0 & \cos(k_{1} - k_{2}) 
\end{array} \right), 
\end{eqnarray}
where $k_{1}$ and $k_{2}$ denote
the x and y components of
the momentum in the plane. 

As detailed earlier, 
for all $K<K_{c}$, a symmetry
operation may be performed to
map any spin model to onto a 
higher $(D>1)$ dimensional one, 
by merely replacing a one dimensional 
momentum
coordinate on a spin chain, $k$, by a linear
combination $\sum_{j=1}^{D} \hat{M}_{j} k_{j}$
where each of the elements $\{\hat{M}_{j}\}$
is a constant matrix, in the 
argument of the interaction
kernel. This is exactly
what is done here. 
The kernel that appears in 
our two dimensional problem, $v$,
may be obtained from
a one dimensional kernel
with the substitution
$\cos k \to \cos (k_{1} + k_{2} \sigma_{3})$
with $\sigma_{3}$ the Pauli matrix 
to the standard nearest neighbor 
kernel on a spin chain.
This transforms the argument of the cosine, $k$, into 
a two dimensional matrix
and makes the interaction look
two dimensional (involving
both $k_{1}$ and $k_{2}$) while
in reality, the system is
one dimensional spin chain 
in disguise. 

The interesting question
is, of course, whether we 
may see similar dimensional
reductions (exact or
approximate) in higher
dimensional gauge
theories of higher
groups.

\section{Phase interference and a mapping onto
a single spin problem}

We will now map classical high dimensional ($d>1$) spin 
problems onto translationally invariant systems
in one dimension.

For the nearest neighbor 
ferromagnet, the basic idea 
is to make a comparison between the 
the kernels
\begin{eqnarray}
\hat{v}_{d} = -2 \sum_{l=1}^{d} \cos k_{l} \nonumber
\\
\hat{v}_{1} = -2 \sum_{l=1}^{d} \cos c_{l} k_{1}
\label{band}
\end{eqnarray}
in $d$ and in one spatial dimensions respectively. 
Note that the one dimensional
problem has, in general, $d$ 
different harmonics of the single
momentum coordinate $k_{1}$. If the
coefficients $c_{l}$ are integers
equal to $R_{l}$ then the one dimensional 
problem amounts to a ferromagnetic 
chain in which each spin 
interacts with $d$ other
spins at distances $\{R_{l}\}_{l=1}^{d}$
away. When evaluating the loop integrals,
we will find that the partition function/
free energy deviate from their
$d$ dimensional values due to ``interference''
between the various harmonics in 
$\hat{v}_{1}(k_{1})$. If all harmonics
acted independently in the effective one dimensional
problem then the $d$ 
dimensional result would be reproduced.
The advantages of the form $\hat{v}_{1}$ are obvious.
Perhaps the most promising  alley is that of ``incommensurate dimensional
reduction''. If this method may be also 
extended to fermionic systems that it would
suggest that we might examine d-dimensional 
problem described by $\hat{v}_{d}$ by 
a bosonization of the one-dimensional
problem of $\hat{v}_{1}$. A 
solution to the one dimensional problem
posed by $\hat{v}_{1}$ with
mutually incommensurate coefficients $\{c_{l}\}_{l=1}^{d}$
will immediately lead to the 
corresponding $d-$ dimensional 
quantities.

\subsection{Commensurate dimensional reduction} 
\label{comm}

Consider a one dimensional ferromagnetic lattice model 
with a nearest neighbor (or $Range=1$) interaction 
and a $Range=n$ interaction: 
$\hat{v}(k_{1}) = -2  (\cos k_{1}+ \cos nk_{1})$.
The loop integrals 
containing terms of the $\cos k_{1}$ 
origin and terms stemming from 
$\cos nk_{1}$ are independent
to low orders.
Mixed terms can survive only
to orders $(1/T)^{n+1}$
and higher. The delta function
constraints $\sum q_{1} =0$
would generate terms identical
to those of the two 
dimensional nearest neighbor 
model with $\hat{v}(k_{1},k_{2})= -2 (\cos k_{1} + \cos k_{2})$
with the two delta function constraints for
the two separate components of the momentum, 
$\sum q_{1}=0$ and $\sum q_{2}=0$, at 
every vertex. Thus, in a sense, this
system is two dimensional
up to $T^{-n}$. Similarly,
if a one dimensional 
chain has interactions
of length one lattice
constant, $R=n$ and $R=m$
i.e.
\begin{equation}
\hat{v}(k_{1}) = - 2  (\cos k_{1} + \cos n k_{1} + \cos m k_{1})
\label{3-d}
\end{equation}
then this system, as fleshed out
by a $1/T$ expansion for 
the partition function
or for the free energy, is three 
dimensional up to $order ~ = ~ n$
if $m=n^2$.  (If
$m=n+1$ then the
triangular ferromagnet 
is generated). Similar, higher
dimensional extensions
similarly follow from 
the lack of commensurability
of the cosine arguments
(for four dimensions  
\begin{equation}
\hat{v}(k_{1}) = - 2  (\cos k_{1} + \cos n k_{1} + \cos m k_{1} + \cos s k_{1})
\end{equation}
with $s= n^3$ etc.)

The corresponding transfer
matrices are trivial to 
write down. For the three dimensional case:
\begin{eqnarray}
\langle s_{1} ... s_{n^2} | T| s^{\prime}_{1} ... s^{\prime}_{n^2}
\rangle = \exp[\beta\{ \frac{1}{2}(s_{1}s_{2}+ ...+s_{n^2-1}s_{n^2}) \nonumber
\\ +\frac{1}{2} (s^{\prime}_{1}s^{\prime}_{2}+ 
...+s^{\prime}_{n^2-1}s^{\prime}_{n^2}) \nonumber
\\ +
\frac{1}{2} (s_{1}s_{n+1}+s_{2}s_{n+2}+ ...+s_{n^2-n}s_{n^2})
\nonumber
\\ + \frac{1}{2} (s^{\prime}_{1}s^{\prime}_{n+1}+s^{\prime}_{2}
s^{\prime}_{n+2}+ ...+s^{\prime}_{n^2-n}s^{\prime}_{n^2}) \nonumber
\\ + s_{n^{2}}s^{\prime}_{1} + (s_{n^{2}-n+1}s^{\prime}_{1} +
s_{n^{2}-n+2}s^{\prime}_{2} + ... + s_{n^{2}}s^{\prime}_{n})
\nonumber
\\ +
(s_{1}s^{\prime}_{1}+ ...+s_{n^2}s^{\prime}_{n^2}) \}] \nonumber
\\ \equiv \exp[\beta {\cal{T}}_{ji}].
\label{transfer}
\end{eqnarray}
In higher dimensional generalizations, ${\cal{T}}_{ji}$ will
be slightly more nested  with the span
$n^{2}$ replaced by $n^{d-1}$.
In Eqn.(\ref{transfer}), the matrix indices $i$ and $j$  
are written in binary 
numerals in terms of the 
$n^{2}$ spins
\begin{eqnarray}
j = \sum_{\alpha=1}^{n^{2}} (s_{\alpha}+1) 2^{\alpha-2}
\end{eqnarray}
where $\alpha$ is the one dimensional coordinate
along the bra spin indices.  
By Eqn.(\ref{transfer}), 
 ${\cal{T}}_{ji}$ are expressed in 
terms of a sum over two digit products where the
digits are those that appear in the binary 
(length $n$ spin) representation of 
the coordinates $j$ and $i$.

As usual, the partition function 
\begin{eqnarray}
Z = Tr[T^{N}] = \sum_{i} \lambda_{i}^{N}
\end{eqnarray}
where $\{ \lambda_{i} \}$ are
the eigenvalues of the transfer
matrix. In the following $\lambda_{\max}$ 
will denote the largest 
eigenvalue.

Note that the transfer matrix
eigenvalues correspond to 
periodic boundary conditions
as strictly required: 
we employed
{\em translational invariance}
to write the partition 
function expansion in 
Fourier space. If the 
boundary conditions are
not periodic then all 
of this is 
void.

By the lack of interference 
effects, the connected diagrams
for the free energy are  
correct to ${\cal{O}}(\beta^{n})$ 
for interactions with $m=n^{2}$
(Eqn.\ref{3-d}).
This along with
\begin{eqnarray}
\beta f =
\sum_{p} f_{p} \beta^{p} \equiv - \frac{\ln Z}{N} = - \ln \lambda_{\max}
\end{eqnarray}
(where the last equality holds in the limit of large 
system size $N$)
and a canonical power expansion
having an infinite radius of convergence
\begin{eqnarray}
\lambda_{\max} = \exp[- \sum_{p} f_{p} \beta^{p}]
= \sum_{m=0}^{\infty} \Big( \frac{1}{m!} (- \sum_{p} f_{p}
\beta^{p})^{m} \Big) \nonumber
\\ = \sum_{p} b_{p} \beta^{p}
\label{eig}
\end{eqnarray}
imply that $\lambda_{\max}$ is correct to ${\cal{O}}(\beta^{n})$.

If we
already know the lower
order coefficients 
corresponding to ${\cal{O}}(\beta^{p})$
with $p=0,1,2,3,...,(n-1)$
when we consider the 
$Range= n$ problem. 
then we may set out to compute is
the coefficient of $\beta^{n}$. 
This will amount to extracting
only a single unknown (i.e. the term corresponding
to the coefficient of $\beta^{n}$) from 
a transfer matrix eigenvalue equation,
This can be done to the next
order etc. recursively.
A simple yet
very long single 
linear relation
gives the largest transfer 
matrix eigenvalue to each 
higher order in the 
inverse temperature
$\beta$.

Explicitly, we
can write longhand
\begin{eqnarray}
\det (T- \lambda) = \epsilon_{i_{1}...i_{2^{n^{2}}}}
\prod_{j=1}^{2^{n^{2}}}
[T_{ji_{j}}(\beta)- \lambda(\beta) \delta_{j i_{j}}]= 0
\label{det}
\end{eqnarray}

All one has to do is
to yank out the ${\cal{O}}(\beta^{n})$
term from the sum over products of 
$2^{n^{2}}$ terms. If we pull out a power $p_{j i_{j}}$ 
from each of the elements $[T_{ji_{j}}(\beta)- 
\lambda(\beta) \delta_{j i_{j}}]$ then those will 
need to satisfy 
\begin{eqnarray}
\sum_{j=1}^{2^{n^{2}}} p_{ji_{j}} = n.
\label{powercounting}
\end{eqnarray}


In each element of the transfer
matrix, $\exp[\beta {\cal{T}}_{ji}]$, 
the coefficient of $\beta^{p}$ is
trivially ${\cal{T}}_{ji}^{p}/p!$. 
For each permutation (or ``path'' of coordinates 
$(i,j)$ within the matrix) to be summed 
for the evaluation of the 
determinant

\begin{eqnarray}
[\sum_{path} {\cal{T}}_{ji_{j}}]^{n} = 
\sum_{p_{1},p_{2}, ...., p_{2^{n^{2}}}} \frac{n!}{p_{1}! p_{2}!
... p_{2^{n^{2}}}!} \prod_{\sum p_{ji_{j}} = n}{\cal{T}}_{ji_{j}}^{p_{j}}
\end{eqnarray}
is exactly the product of the coefficients of $\beta^{p_{ij}}$
such that those lead a net power $\beta^{n}$ (i.e. satisfying 
Eqn.(\ref{powercounting}))
from that particular path.

It follows that net contributions 
stemming from $T_{ji_{j}}(\beta)$  in 
Eqn.(\ref{det}) 
are 
\begin{eqnarray}
F^{(n)} \equiv  \frac{1}{n!} \sum_{all ~ paths} \epsilon_{path} [\sum_{(ji_{j}) ~ in ~a ~
given~ path} {\cal{T}}_{ji_{j}}]^{n}
\end{eqnarray}
where $\epsilon_{path}$ simply denotes 
the sign of the given permutation $\{ j \}_{j=1}^{2^{n^{2}}} \rightarrow \{ i_{j}\}$.

To take into account the 
products including 
$\lambda(\beta)$ let us
define  
\begin{eqnarray}
D^{(n;p)}_{2m} = \frac{1}{(n-p)!} 
\sum_{all~ 2m~ paths}  \nonumber
\\  \epsilon_{path} 
[\sum_{(ji_{j}) ~ in ~a ~given~ path} ^{\prime}
{\cal{T}}_{ji_{j}}]^{n-p}
\end{eqnarray}

where the summation is over paths 
going through all possible $(2m)$ 
given points on the diagonal and $\sum^{\prime}$
denotes a summation over ${\cal{T}}_{ji_{j}}$ in the
given paths sans the contributions from
the $(2m)$ diagonal points. 
The determinant paths in $\sum_{all~ 2m~ paths}$ 
can pass through more then $(2m)$ diagonal points-- it is
just that we need to sum over all
those that in their pass also traverse
all possible sets of $(2m)$ given points 
on the diagonal and where those points 
are excluded from the second
${\cal{T}}_{ji_{j}}$
summation.

The $\beta^{n}$ component of Eqn.(\ref{det})
is

\begin{eqnarray}
F^{(n)} + \sum_{p=0}^{n} \sum_{m=1}^{2^{n^{2}-1}}   D_{m}^{(n;p)}
\sum_{\sum_{i=1}^{2m} p_{i} =p} b_{p_{1}} ... b_{p_{2m}} = 0 
\label{rec}
\end{eqnarray}


where $\{b_{p}\}_{p=0}^{n}$ appear in 
the expansion of the largest 
eigenvalue (Eqn.(\ref{eig})).

If the coefficients $ \{b_{p}\}_{p=0}^{n-1}$
then Eqn.(\ref{rec}) where $b_{n}$ appears 
there only linearly gives $b_{n}$ and one may then
proceed to look 
at the eigenvalue equation
for the next value of $n$.

In the absence of
an external field,
for $n=1$ the 
largest eigenvalue for the 
$d$ dimensional problem is 
$\lambda_{+} = 2 \cosh d \beta$
and other eigenvalue
$\lambda_{-} = 2 \sinh d \beta$.
For the three dimensional 
case setting $m=n^{2}=n=1$ 
leads to the one dimensional momentum
space kernel $\hat{v}(k_{1}) = - 6 \cos k_{1}$
which in real space corresponds to 
\begin{eqnarray}
H = - 3 \sum_{i} S_{i} S_{i+1}.
\end{eqnarray}
Of the two eigenvalues only $\lambda_{+}$ has a nonzero $b_{0}$. 
Knowing the values $b_{0}=2$ and $b_{1}=0$ of the largest 
eigenvalue we may
proceed to find $b_{2}$ from 
the Eqn.(\ref{rec}) when $n=2$
etc. This, of course, may be extended
to systems with magnetic
fields.

Setting $b_{0}=2$ in 
Eqn.(\ref{rec}) we find 
\begin{eqnarray}
- b_{n} \sum_{m=1}^{2^{n^{2}-1}}  m 4^{m} D_{m}^{(n;n)} 
  = \nonumber
\\ 
F^{(n)}  + 
\sum_{p=0}^{n-1} 
 \sum_{~p= p_{1}+p_{2}} b_{p_{1}} b_{p_{2}}
D_{1}^{(n;p)} \nonumber
\\ + 
\sum_{p=0}^{n} \sum_{m=2}^{2^{n^{2}-1}}   D_{m}^{(n;p)}
\sum_{\sum_{i=1}^{2m} p_{i} =p~~ ~with~ p_{i} \neq p }
b_{p_{1}} ... b_{p_{n}}. 
\end{eqnarray}
This is the rather explicit recursive relation
giving $b_{n}$ once $\{b_{p}\}_{p=0}^{n-1}$
are known. To find out the coefficient of 
$b_{n}$ let us look at

\begin{eqnarray}
\overline{D}_{s}^{(n;n)} = \sum_{exactly~2s~diagonal~paths} \epsilon_{path},
\end{eqnarray}

where the summation admits only those
permutations that lead to
exactly $2s$ diagonal elements. For the $2^{n^{2}}$ diagonal path 
in the matrix ${\cal{T}}$: $\epsilon_{path}=1$.
For all paths threading $2^{n^{2}-2}$ diagonal
sites $\epsilon_{path}=-1$ and the number
of such paths is ${2^{n^{2}} \choose 2}$.
For the general case the last quantity amounts
to 
\begin{eqnarray}
\overline{D}_{s}^{(n;n)} = (-1)^{s} {2^{n^{2}} \choose 2s}.
\end{eqnarray}
Evaluating, we find
\begin{eqnarray}
D_{m}^{(n;n)} = \sum_{s=m}^{2^{n^{2}-1}} \overline{D}_{s}^{(n;n)}
= 
\frac{(2^{n^{2}})!}{(4m)!(2^{n^{2}}-4m)!}\nonumber
\\ \times 
F_{pq}[\{1,\frac{m}{2}-2^{n^{2}-1},\frac{1}{2}+\frac{m}{2}-2^{n^{2}-1}\},
\nonumber
\\ \{\frac{1}{2}
+ \frac{m}{2},1+2m\},1]
\end{eqnarray}
where $F_{pq}$ is a generalized hypergeometric function.

\begin{figure}
\begin{center}
 \epsfig{figure=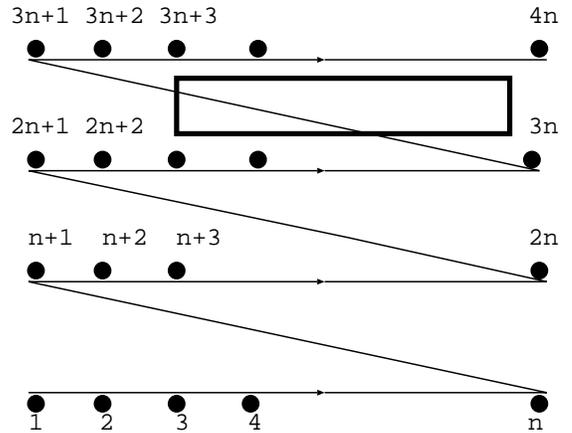,width=0.85\linewidth}
\end{center}
\caption[]{Lacing of an $(n \times L$) 
slab (with $L \to \infty$). All real space 
high temperature diagrams up to order $(\tanh^{n} \beta$) 
such as the closed rectangle shown 
above are of equal value in 
both this one dimensional 
system with interaction ranges
$R=1,n$ 
and the full $(  L \times L$) two dimensional
model with nearest neighbor interactions.
The width of the slab ``$n$'' serves 
as an inverse temperature axis in the following
sense- the wider it is, the higher order in 
$1/T$ that we may advance towards 
the full two dimensional model.
We may get an analogous ``three-dimensional''
slab with ``two imaginary time axis''
if we set $m=n^{2}$. The path will
then thread $n^{2}$ sites before 
continuing upward.}
\label{string}
\end{figure}

Insofar as simple geometric visualization 
is concerned, it is amusing to
note, as shown in Fig.(\ref{string}) for
the two dimensional case, that employing 
the usual high temperature
expansion in powers of $\tanh \beta$ one
would reach the same conclusion
regarding the correctness (up to order $n$) 
of the partition function evaluated for 
a width $n$ slab vis a vis the 
partition function of the two 
dimensional system. We 
may look at the a finite thickness $(n$) slab 
of the two dimensional lattice along which we
apply periodic boundary conditions.
Let us now draw a string going 
along one row of length $n$,
after which it would jump to the next row, scan
it for $n$ sites, jump to the next one, and so
on. On the one dimensional laced string,
the system is translationally invariant and
the interactions are of ranges $R=1,n$.
Or, explicitly, by counting the number of 
closed loops in real space
(employing the standard, slightly
different, diagrammatic expansion
in powers of $\tanh \beta$),
we see that the terms
in this expansion are also identical
up to order $\tanh^{n} \beta$ \cite{Triangular}.

A lacing of a two dimensional slab by a one 
dimensional string, as shown in 
the figure, is one of the backbones  
of Density Matrix Renormalization
Group Theory when applied
to two dimensional problems.

\subsection{Incommensurate dimensional reduction} 
\label{incomm}

\bigskip

 The one dimensional continuum limit ($- \Lambda < k_{1} < \Lambda$
with $\Lambda \to \infty$)  
momentum space kernel
\begin{equation}
\hat{v}(k_{1})= - 2 \sum_{l=1}^{d} \cos c_{l} k_{1}
\label{soo}
\end{equation}
with  mutually incommensurate 
$\{c_{l}\}$ will 
reproduce the expansion 
with the kernel in Eqn.(\ref{d-dim})
for the  $d$ dimensional hypercubic lattice system \cite{err}.

\bigskip

The proof is quite easy. The most general integral is 
of the form 
\begin{eqnarray}
 \Big[
\int_{-\Lambda}^{\Lambda}...
\int_{-\Lambda}^{\Lambda}
\prod_{b=1}^{\mbox{loops}} \frac{dk_{b}}{2 \Lambda} \nonumber
\\ ~ ~
\prod_{a=1}^{\mbox{propagators}} \sum_{l=1}^{d}   
\exp[i \sum_{b=1}^{\mbox{loops}} M_{ab} k_{b} c_{l}]~ \Big]
\label{Ll}
\end{eqnarray}
where the index $b$ runs over the various 
independent loop momenta
($k$ is still merely a 
scalar for this one dimensional
problem). Unless, for a given $k_{b}$ integration,
the argument of the exponent is identically
zero (i.e. corresponding to
a term that would be generated 
in the d-dimensional nearest neighbor
problem) an ``interference term'' results.
However, such a term is down by ${\cal{O}}(\Lambda^{-1})$ 
by comparison to the ``good'' noninterference 
terms that occur in the d-dimensional
problem. The 
canonical loop integral reads
\begin{equation}
\frac{1}{2 \Lambda} \int_{-\Lambda}^{\Lambda} dk_{b} \exp[ic k_{b}] \rightarrow
\frac{\pi \delta(c)}{\Lambda}
\label{las}
\end{equation}
where $c$ is a linear combination
of the $d$ coefficients $\{c_{l}\}$.
Unless $c$ vanishes identically
(i.e. unless a ``good'' non-interference term occurs),
then $c \neq 0$ if $\{c_{l}\}$ are chosen to
be mutually incommensurate.
Whenever $c=0$, the integral in Eqn.(\ref{las}) 
transforms into the Kronecker delta $\delta(c,0) =1$
in the limit $\Lambda \to \infty$ and the 
integral attains exactly 
the same value that it 
would take on for the $d-$dimensional
hypercubic lattice system.
Similarly, in this limit 
all ``bad'' interference 
terms will evaporate
in all diagrams.

As this holds for any diagram, 
this conclusion is also
valid for any set of 
diagrams (not only those
encountered in computation 
of the partition function
or the free energy),
e.g. $C_{V}$, $\chi$ etc.

As a consequence we may
state that for a single 
spin 1/2 particle
with an action 
\begin{eqnarray}
S = \int ~ d \omega  ~ \overline{\psi}(-\omega) ~\hat{v}(\omega) ~\psi(\omega)
\end{eqnarray}
where $\psi(\tau)$ is the two component
spinor, the partition function
is identically the same as that of 
the $d$ dimensional nearest
neighbor ferromagnet.

The proof of this
statement trivially
follows from breaking up
the $[0,\beta]$ segment
along the imaginary time axis
into $L$ pieces and allowing 
$L \rightarrow \infty$. 

We have mapped the entire three
dimensional Ising model onto
a single spin 1/2 quantum particle \cite{EQU}.

The Fourier transform of Eqn.(\ref{soo})
on a lattice reads
\begin{eqnarray}
\hat{V}(x_{1}) = - 2  \sum_{l=1}^{d} 
[(c_{l}+x_{1}) \sin \pi(c_{l}-x_{1}) \nonumber
\\
+ (c_{l}-x_{1})] \sin
\pi(x_{1}+c_{l})]\nonumber
\\ \times [(c_{l}-x_{1})(c_{l}+x_{1})]^{-1}, 
\label{redu}
\end{eqnarray}
a sum of shifted Coulomb like interactions. 
If an average $(2 \Lambda)^{-1} \int_{-\Lambda}^{\Lambda} dc_{l}$ is performed over 
each of the $d$ coefficients $\{ c_{l}\}$ 
then the resulting quantities 
are the exact $d$-dimensional ones. 
If ``bad'' interference terms
result then their value is $O(1/c)$.
In the averages, it is the large coefficients that dominate and these
will lead to a vanishing average value of 
all interference terms.

\section{Incommensurate Dimensional Reductions 
of Quantum Models}

The incommensurate dimensional reduction
averaging theorem which 
we just proved by interchanging the order of integrations over
$\{c_{l}\}$ and the loop momenta $\{\vec{k}_{b}\}$ holds for 
any translationally invariant system (spin, bosonic, fermionic)
in which the unperturbed action $S_{0}$ 
links two nearest neighbor sites. 
For a $(d+1)$ dimensional quantum
lattice system, the Green's function conjugate to 
$S_{0}$ in the undualized (non Hubbard-Stratonovich
transformed) system 
\begin{eqnarray}
G_{0} = \Big[\beta \Big( - 2 \sum_{l=1}^{d+1}  \cos k_{l} + A \Big)\Big]^{-1}.
\label{QUT}
\end{eqnarray}

As before, if the magnitude of the squared mass $|A| =const > 
2(d+1)$ then a  geometric series expansion 
in A$^{-1}$ may be performed
to generate a sum of harmonics.
For each given single diagrammatic term,
our previous
proof may be applied. 

We note that fermionic electron (or other) fields might, perhaps, be
formally bosonized on each individual chain.
The one dimensional band,  
\begin{eqnarray}
\hat{v}_{1} = -2 \sum_{l=1}^{d} \cos c_{l}
k_{1},
\label{one-d-}
\end{eqnarray}
is now a simple sum of $d$ cosines (au lieu of the 
standard single tight binding cosine). For each given 
set $\{c_{l}\}_{l=1}^{d}$, the band dispersion of Eqn.(\ref{one-d-}) 
may be easily linearized about 
its respective Fermi points. Consequently 
the standard bosonization methodology 
may be applied. In the large $\{c_{l}\}$ limit,
the number of Fermi points becomes infinite. 
A one dimensional Fermi system with 
$(2n)$ Fermi points is expected to 
be equivalent to a d-dimensional
system up to $O(\beta^{n})$.

The astute reader will note 
that we may just as well use 
Eqn.(\ref{QUT}) to study a 
classical spin system (with 
the upper bound $d+1$ 
replaced by $d$) and 
prove everything that   
we have previously stated
without the need to
consider the dual model.
Such an approach is
not as rigorous as
that followed
hitherto. First, the
geometric series expansion
in the harmonics $\exp[i k_{l}]$ 
converges only for a sufficiently
absolute value of the mass-
or, equivalently for temperatures
far enough from the mean field
temperature. Second, the expansion
in $G_{0}$ is in general explicitly
asymptotic as the perturbing piece is 
of higher order in the fields 
than $H_{0}$. The dual formulation
was free of these pathologies.

\section{Permutational symmetry}
\label{permutational}

The classical spin spherical model (or  $O(n \rightarrow \infty)$) 
partition function
\begin{equation}
  Z = const \left(\prod_{\vec{k}}
    \left[\frac{1}{\sqrt{\beta[\hat{v}(\vec{k})+\mu]}}\right]\right),
\end{equation}
where the chemical
potential
$\mu$ satisfies
\begin{eqnarray}
\beta = \int \frac{d^{d}k}{(2 \pi)^{d}} \frac{1}{\hat{v}(\vec{k}) + \mu},
\end{eqnarray}
is invariant under permutations of $\{\hat{v} ( \vec{k})\} \rightarrow
\{\hat{v}(P\vec{k})\}$. In the above, the permutations
\begin{eqnarray}
\{\vec{k}_{i} \}_{i=1}^{N} \rightarrow \{ P \vec{k} \}
\end{eqnarray}
correspond to all possible shufflings of the 
$N$ wavevectors $\vec{k}_{i}$. Although quite simple,
this is not universally realized. Several authors
have attempted to compute the critical exponents
in the spherical limit (via an RG calculation)
for systems having different minimizing manifolds
yet all sharing the same relevant density of states.
This quest was not very economical. As unrealized by 
these authors, by permutational symmetry these models are identical.

This simple invariance allows all d-dimensional 
translationally invariant systems to 
be mapped onto a 1-dimensional one. Let us design an effective
one dimensional kernel $V_{eff}(k)$ by
\begin{equation}
  \int \delta[\hat{v}(\vec{k})-v] d^{d}k =
  |\frac{dV_{eff}}{dk}|_{V_{eff}(k)=\hat{v}}^{-1}.
\end{equation}
The last relation secures that
the density of states and consequently
the partition function is preserved.  For 
the two-dimensional nearest-neighbor ferromagnet:
\begin{eqnarray}
  |\frac{dV_{eff}}{dk}|^{-1} =\rho(V_{eff}) \nonumber
\\ = c_{1}
  \int_{0}^{1}\frac{dx}{\sqrt{1-x^{2}}\sqrt{1-(V_{eff}+x-2)^{2}}},
\end{eqnarray}
and consequently 
\begin{eqnarray}
  k(V_{eff})= c_{1}\int_{0}^{V_{eff}}
  F(\sin^{-1}\sqrt{\frac{2}{(3-u)u}},\nonumber
\\ \frac{\sqrt{4u- u^{2}}}{2}) du,
\label{elliptic}
\end{eqnarray}
where $F(t,s)$ is an incomplete elliptic integral of the first kind.
Eqn.(\ref{elliptic}) may be inverted
and Fourier transformed to find the effective one dimensional 
real space kernel $\hat{V}_{eff}(x)$.
We have just mapped the two dimensional nearest 
neighbor ferromagnet onto a one dimensional 
system.
In a similar fashion, within the spherical 
(or equivalently the $O(n \rightarrow \infty)$) limit
 all high dimensional problems may be mapped
onto a translationally invariant one dimensional 
problem. It follows that 
the, large $n$, critical exponents of the
$d$ dimensional nearest neighbor ferromagnet
are the same as those of translationally invariant 
one dimensional system with longer range
interactions. We have just shown 
that a two dimensional $O(n \gg 1)$ 
system may has the same thermodynamics
as a one dimensional system.
By permutational symmetry,
such a mapping may be performed
for all systems irrespective
of the dimensionality of the 
lattice or of the nature of the 
interaction (so long
as it translationally
invariant). This demonstrates once again
that the notion of universality 
(with dependence only on the order parameter symmetry, 
dimensionality etc.) may apply only to the canonical 
interactions.

The lowest order term
breaking permutational symmetry in our high temperature
expansion is $\eta^4(\vec{x})\eta^{4}(\vec{y})$.  
Thus permutational symmetry is broken 
to ${\cal{O}}(\beta^{4})$ for finite $n$.
For a constraining term (e.g. $\sum_{\vec{x}} \ln[\cosh[\eta(\vec{x})]]$ for
$O(1)$ spins) symmetric in $\{\eta(\vec{k})\}$ to a given order, one
may re-arrange the non-constraining term $\sum_{\vec{k}}
\hat{v}^{-1}(\vec{k}) |\eta(\vec{k})|^{2} = 
\sum_{\vec{k}} \hat{v}^{-1}(P \vec{k})
|\eta(P \vec{k})|^{2}$) and relabel the dummy integration variables
$H[\{\eta(\vec{k})\}] \rightarrow H[\{\eta(P^{-1}\vec{k})\}]$ to effect
the constraining term augmented to a shuffled spectra $\hat{v}
(P \vec{k})$. \cite{perma}

{\bf Acknowledgments.}

This research was supported by the Foundation
of Fundamental Research on Matter (FOM), which is sponsored by the
Netherlands Organization of Pure research (NWO). 
It is a pleasure to thank E.~ Altman, ~A.~ Auerbach, 
~A.~ V.~ Balatsky, ~S.~ A.~ Kivelson, 
~W.~ van Saarloos, and ~J.~ Zaanen for discussions.

\section{Appendix A: The Hubbard Stratonovich
Transformation for $O(n)$ spin systems}
\label{apa}

For the benefit of our uninitiated 
readers we present the standard 
Hubbard Stratonovich transformation.
This ``transformation'' merely relies on the elementary Gaussian identity
\begin{equation}
\int_{-\infty}^{\infty} \frac{d \eta}{\sqrt{2 \pi}} \exp[ -\frac{1}{2} K \eta^{2} + \phi \eta] = K^{-1/2} \exp[ \frac{\phi^{2}}{2K}].
\end{equation}
As well known, this easily generalizes (upon diagonalization) to
\begin{eqnarray}
\int (D \eta) \exp[-\frac{1}{2} (\eta, K \eta) +(\eta,\phi)] \nonumber
\\ = (\det K)^{-1/2} \exp \frac{1}{2} (\phi, K^{-1} \phi).
\end{eqnarray}

For our particular purposes, note that
\begin{eqnarray}
\sqrt{\det(2 \pi (-V^{-1}))} = \int D \eta~
\exp[-\frac{1}{2}(sV-\eta)\nonumber
\\ (-V^{-1})
(Vs-\eta)]. 
\end{eqnarray}m

Thus 

\begin{eqnarray}
Z = \sum_{s} \exp(-\frac{1}{2} s,Vs) \nonumber
\\ = \sqrt{ \det \Big({\frac{-2 V}{\pi} \Big)}} \int D \eta~
\exp[ \frac{1}{2} \eta, V^{-1} \eta] 
\nonumber
\\ \sum_{s} \exp[-s, \eta] \nonumber
\\ = \sqrt{\det \Big( \frac{-2 V}{\pi} \Big)} \int D \eta \exp[-\tilde{H}]
\end{eqnarray}
where 
\begin{eqnarray}
\tilde{H} = -\sum_{i} \ln(\cosh \eta_{i}) -\frac{1}{2} (\eta, V^{-1}
\eta).
\label{last}
\end{eqnarray}

In the above $s=(S_{1},S_{2},...,S_{N})$ where 
$i$ label the sites $\vec{x}$. For an Ising system $S(\vec{x}) = \pm 1$.
In Eqn.(\ref{last}), the relation
\begin{eqnarray}
\sum_{(s)} \exp(- s,\eta) = \sum_{(s)} \exp(- s_{1} \eta_{1}-...-s_{N}
\eta_{N}) \nonumber
\\= \prod_{i=1}^{N} (2 \cosh \eta_{i})
\end{eqnarray}
was employed.

In an $O(n)$ spin system,
tracing out over the spins
leads to
\begin{eqnarray}
\sum_{(s)} \exp(- s,\eta) = \prod_{i=1}^{N}\Big[\int 
d \Omega_{S_{i}} \exp(-\vec{S}_{i} \cdot \vec{\eta}_{i}) \Big] \nonumber
\\ = \prod_{i=1}^{N} \Big[\Gamma(n/2)~
  (\frac{2}{|\vec{\eta}_{i}|})^{n/2-1}~
  I_{n/2-1}(|\vec{\eta}_{i})|)\Big].
\end{eqnarray}

\section{Appendix B: The Eight-fold ``Porcupine'' 
Spin Model and other ``Platonic'' Spin Models}
\label{apb}

For ``$Z_{2} \otimes Z_{2} \otimes Z_{2}$''
models, the three-dimensional spins explicitly read

\begin{equation}
{\bf s} = (\sigma_{1},\sigma_{2},\sigma_{3})
\end{equation}
with 
\begin{equation}
\sigma_{i}= \pm 1
\end{equation}
(i.e. spin $\frac{1}{2}$ like).
This corresponds to the 8 possible states (the vertices of the unit cube)
($\pm 1, \pm 1, \pm 1$), or in a more
pompous notation- the eight 
direct product states of three spin-1/2 particles 
$| \sigma_{1} \rangle  \otimes |\sigma_{2} \rangle \otimes 
|\sigma_{3} \rangle$ states. For most canonical rotationally and translationally 
invariant models $\langle \sigma_{1} \sigma_{2} \sigma_{3 }| H | \sigma_{1}
\sigma_{2} \sigma_{3} \rangle$ decomposes into a 
sum of three decoupled Hamiltonians each living 
within its own individual two dimensional ($|\sigma_{i} \rangle$) 
subspace. The three dimensional ``single
particle (spin)'' problem will reduce to 
a sum of the three individual non-interacting 
Hamiltonians each corresponding to a single free  
(spin-1/2)  ``particle'' having only two states.
A Hamiltonian containing only $[{\bf S_{i}} \cdot {\bf S_{j}}] $ terms 
(with ${\bf i}$ and ${\bf j}$ lattice sites),
amounts to 
\begin{equation}
H(\{{\bf S_{i}}\}) = \sum_{{\bf x_{i}}, {\bf{x_{j}}}}~~
~V({\bf x_{i},\bf x_{j}}) [{\bf{S}_{i}} \cdot {\bf{S}_{j}}] 
= \sum_{\alpha=1}^{3} {\cal{H}}(\{\sigma^{\alpha}_{i}\})
\end{equation}
where ${\cal{H}}$ are Ising Hamiltonians, i.e.
\begin{equation}
{\cal{H}}(\{\sigma^{\alpha}_{i}\}) = \sum_{{\bf x_{1}}, {\bf{x_{2}}}}~~\sigma^{\alpha}_{1}
~V({\bf x_{1},\bf x_{2}}) ~\sigma^{\alpha}_{2}.
\end{equation}
The partition function
\begin{equation}
Z = \prod_{\alpha=1}^{3} Z_{\alpha}= (Z_{Ising})^{3}.
\label{factorization}
\end{equation}
The $Z_{2}^{3}$ model 
behaves exactly like an Ising model
(or a $q=2$ Potts model).
This may be extended to other $\tilde{N}$
state clock models.
When $\tilde{N}=2$ the ``clock model'' 
is trivially the $q=2$ Potts 
(Ising) model. When $\tilde{N}=n+1$ in the 
$n-$dimensional ``clock model'', 
a mapping to an $(n+1)$ state Potts
model is possible (by constructing
a regular $n$ dimensional tetrahedra).
For $\tilde{N}=2^{n}$ (the $n$ dimensional
hypercube), the model
may be mapped (once again)
onto $n$ decoupled $q=2$ Potts 
models. In two dimensions, $q_{c}=4$ is the
critical number of 
states required for the nature
of the transition to vary
in the unfrustrated ferromagnetic
Potts model. Thus,
in the unfrustrated
case, as $\tilde{N}$ is monotonically
increased (for $n>3$), 
the nature of the transition
changes, at least 
twice, from that with  
($\tilde{N}=2;  ~ 2 = q  \le q_{c})$ to ($\tilde{N}= q = n+1 >q_{c}$) 
to the $ q\le q_{c}$ case ($\tilde{N}=2^{n}; ~ 2= q \le q_{c}$) 
once again. For all $\tilde{N} < n+1$  uniformly spaced polarizations, 
with overall isotropic orientation, the model is
exactly equivalent to a $q=\tilde{N}$ state Potts model;
in this case, the discrete polarizations 
span an $N-$dimensional subspace
embedded in the full $n-$dimensional
spin space.

The three-dimensional ``clock model'' is restricted only
to very special $\tilde{N}$ values corresponding 
to the perfect solids (the six perfect polyhedra
-- the ``Platonic solids''-
when $n=3$) and when $n>3$, 
the uniformly spaced spin polarizations 
are restricted to the vertices  
of the (special 
small set of allowed) regular 
polytopes possible for that
value of $n$. The $\tilde{N} \rightarrow 
\infty$ limit may 
not be taken- non-infinite $\tilde{N}$'s
are bounded by a finite number (e.g. $\tilde{N}_{\max}=60$ for $n=3$).

When $\tilde{N} = \infty$ (not the limit)
for a given $n$ the 
continuous $O(n)$ model 
is attained -this model has second
order phase transitions
in $d=3$ for its unfrustrated 
ferromagnetic variant.

When $n=3$,~ the only possible 
finite $\tilde{N}$ variants are 
$\tilde{N}=4$ (the tetrahedron spin model that may be mapped
onto the $q=4$ state Potts model as 
just discussed), $\tilde{N}=8$
(the cube or hexahedron -~ $q_{eff}=2$ as discussed 
in on page 1), $\tilde{N}=6$ (the octahedron),
$\tilde{N}= 20$ (the dodecahedron), 
$\tilde{N}=12$ (the icosahedron),
and $\tilde{N}=60$ (the soccer-ball
or truncated icosahedron).

For the $\tilde{N}=6,12,20,$ and 60
variants a mapping into a
model similar to the Potts
model is possible. Here the 
number of allowed individual 
two-spin energy states
is no longer effectively
two (as in the Potts models
for equal and unequal  
spin polarizations
of neighboring spins) but is slightly 
larger - a trivially generalized 
Potts model (e.g. there are three
different possible values for a 
spin-spin interaction 
in the $\tilde{N}=6$ model).


The octahedron ($\tilde{N}=6$) model is equivalent 
to a ``three-color
dilute Ising model'' where
each site is colored
red, white, or blue.
On each individual
color cluster
independent Ising 
spins interact
amongst themselves.
Consequently, a sum over all possible 
colorings (dilute inter penetrating
color sites) of the lattice 
sites is to be 
performed.

{\bf References}\\

\begin{eqnarray}
\vec{S}(\vec{x}) = \frac{1}{N} \sum_{\vec{k}} \vec{s}(\vec{k}) \exp[i
\vec{k} \cdot \vec{x}]
\label{Fr}
\end{eqnarray}
then the normalization constraint at each
site $\vec{x}$ reads:
\begin{eqnarray}
1 = S^{2}(\vec{x}) = \frac{1}{N^{2}} \sum_{\vec{k},
\vec{k}^{\prime}}  \vec{s}(\vec{k}) \cdot \vec{s}(\vec{k^{\prime}})
\exp[i(\vec{k}+ \vec{k}^{\prime}) \cdot \vec{x}].
\label{CONstraints}
\end{eqnarray} 

A permutation of $\{ \vec{s}(\vec{k}) \}$
(which is equivalent to a permutation
of the spectra $\{v(\vec{k})\}$ 
within the quadratic interaction
kernel) would transform the $N$ 
constrains of Eqn.(\ref{CONstraints}) 
into another set of 
equations. 

The sole normalization
constraint which
is permutationally
symmetric is the 
global normalization 
constraint (appropriate 
to the $O(n \to \infty)$ 
spherical model)
\begin{eqnarray}
1 = \frac{1}{N^{2}} \sum_{\vec{k},
\vec{k}^{\prime}}  \vec{s}(\vec{k}) \cdot \vec{s}(\vec{k^{\prime}}).
\end{eqnarray} 
This illustrates why permutational symmetry 
must trivially hold, to all orders, in the $O(n \to \infty)$ 
limit (and only in this limit).

For the quantum spin system, 
\begin{eqnarray}
\vec{s}(\vec{k}) \times \vec{s}(\vec{k^{\prime}}) = i
N \vec{s}_{\vec{k}+\vec{k}^{\prime}}.
\label{spin-com}
\end{eqnarray}
Just as with the normalization 
conditions for the classical
spin models, these commutation relations
are not invariant under permutations
of $\{ \vec{s}(\vec{k}) \}$.

We mention, in passing, 
that permutational symmetry 
trivially holds in all quadratic (non-interacting)
translationally invariant Bose and Fermi systems.
The respective commutation and anticommutation
relations
\begin{eqnarray}
[a^{\dagger}(\vec{k}^{\prime}), a(\vec{k})] = \{ c^{\dagger}(\vec{k}),
c(\vec{k}^{\prime}) \} = \delta_{k,k^{\prime}}
\end{eqnarray}
are invariant under permutations
$ \hat{O}(k) \to \hat{O}(Pk)$ for all
bosonic or fermionic $\hat{O}$.


\begin{thebibliography}{99}
\bibitem{Cool} 
Explicitly, the inverse kernel of the 
nearest neighbor ferromagnet reads
\begin{eqnarray} 
V^{-1}(\vec{x},\vec{y}) = \beta^{-1} \int \frac{d^{d}k}{(2 \pi)^{d}}
~ \frac{\exp[i \vec{k} \cdot \vec{x}]}{A - 2 \sum_{l=1}^{d} \cos k_{l}}
\end{eqnarray}
where $A$ denotes the strength of a possible additional onsite 
$A [\vec{S}(\vec{x})]^{2}$ interaction term which 
is merely a constant ($A$) for the normalized
$O(n)$ spins. For $A=2d$, the denominator is none 
other than the lattice Laplacian which scales as 
$k^{-2}$ for small momenta and whose inverse 
Fourier transform is a real 
space Coulomb kernel. For $A>2d$, a Yukawa like 
screened Coulomb interaction kernel 
(with a screening length $A^{-1/2}$) 
appears.
\bibitem{Blat} The
asymptotically convergent character
of generic $\epsilon$ expansions and 
$\phi^{4}$ field theories was frequently overlooked 
and has led to quite a few incorrect 
conclusions in the past. For instance, 
the exact equivalence
of the diagrammatic expansions
for the uniform and random
magnets in $(d-2)$ and  $d$
dimensions respectively,
has led to the speculation
that the two problems may be 
identical. If correct, this
would have implied that the lower 
critical dimension of the 
random field Ising model would 
be three and not two as follows 
from Imry-Ma arguments. 
The flaw in such conclusions
resting on the equivalence
of two asymptotic diagrammatic series 
was pointed by D. S. Fisher \cite{Fish}.
Our expansion is 
convergent for $\beta < \beta_{c}$
(the inverse critical temperature)
and is explicitly free from many 
of the spurious pathologies present 
in canonical diagrammatic
expansions.
\bibitem{tsitsi}
If a twist $\alpha$ is applied
across the sample (e.g. $\alpha =1$ corresponds
to periodic boundary conditions and 
$\alpha = -1$ to antiperiodic boundary conditions) 
then  $\{ z_{n}\}$, with $z_{n} \equiv \exp[ik_{n}]$,
are roots of the 
equation $z^{L}- \alpha =0$;  
the sum of these polynomial roots
is zero. More generally, Eqn.(\ref{lat-summ})
holds for any system in which 
the sum of the all allowed values,
$\sum_{n} z_{n}$, is equal to zero.
\bibitem{N}
As a technical aside, note that 
on the lattice (unlike the standard
continuum theory) each vertex 
conserves momentum only
up to a reciprocal lattice 
vector $\vec{K} = 2 \pi (n_{1},n_{2}, ..., n_{d})$
with $n_{i}$ integers.
Formally, this is as $\vec{x}$ only take on discrete
values on a lattice,

\begin{equation}
\sum_{\vec{x}} \exp[i \vec{k} \cdot \vec{x}] = N \sum_{\vec{K}}
\delta_{\vec{k},\vec{K}}
\end{equation}
where $\vec{K}$ are reciprocal lattice 
vectors. 
 
Thus at each vertex of order $2m$ we have 
\begin{equation}
\frac{1}{(2 m)!} \sum_{\vec{K}} \delta(\vec{q}_{in}-\vec{K},0)
\end{equation}
where $\vec{q}_{in}$ denotes the net momenta 
flowing into the vertex. As for all lattice 
models the propagators are periodic 
$\hat{v}(\vec{q}) = \hat{v}(\vec{q}+\vec{K})$
for each of the $N$  reciprocal lattice vectors $\vec{K}$.
As such the momentum conservation constraint 
may be replaced
by 
\begin{equation}
\frac{1}{(2 m)!} \sum_{\vec{K}} \delta(\vec{q}_{in}-\vec{K},0) 
\rightarrow N \frac{1}{(2 m)!}  \delta(\vec{q}_{in},0). 
\end{equation} 
In the aftermath, each vertex carries a factor
of the volume ($N$) multiplying the discrete
Kronecker delta function. After the momentum
conservation constraints are taken care of
at each vertex, we will be left
with a summation over all remaining independent loop
momenta $\{ \vec{k}_{b} \}$. Each one of these 
summations is to be replaced by the standard
\begin{equation}
\sum_{\vec{k}} \rightarrow N \int_{-\pi}^{\pi} ...  
\int_{-\pi}^{\pi} \frac{d^{d}k}{(2 \pi)^{d}}. 
\end{equation}
Thus, at the end we will be left with 
a factor of
\begin{equation}
N^{\mbox{loops + vertices - propagator~ lines}}
\end{equation}
multiplying an integral over the independent loop
momenta $\{ \vec{k}_{b} \}$ in Eqn.(\ref{loop}).
Compounding all together, as in each closed diagram
the number of independent loops + number of
delta function momentum conserving vertices  =
1 + number of propagator lines, a net factor of $N$
should be attributed to each independent closed
bubble just as it does in the 
conventional continuum theories.
\bibitem{Gibbs} 
As noted earlier, for any
lattice model 
\begin{equation}
\hat{v}(\vec{k}) = \hat{v}(\vec{k}+\vec{K}),
\end{equation}
as $\exp[i \vec{K} \cdot \vec{x}]=1$ in 
\begin{eqnarray}
\hat{v}(\vec{k})= \sum_{\vec{x}} \hat{V}(\vec{x}) 
\exp[i \vec{k} \cdot \vec{x}].
\end{eqnarray}
The sum over $\vec{x}$ spans all
discrete lattice sites. 


The way to analytically continue
$v(\vec{k})$ when $k_{1}$ becomes complex is 
to make the $k_{2}$ dependence also 
invariant under shifts 
by $(2 \pi)$. Within each Brillouin
Zone the $k_{2}$ dependence stems
from the analytic continuation
of the functional 
dependence on 
the real part.
The only complication
is the Gibbs' phenomenon
in $k_{3}$ which occurs 
when taking the Fourier 
transform of the lattice 
$\hat{V}(\vec{x})$.
This is slightly 
avoided for the real
part of $V(\vec{x})$ as follows:
Under parity  and
all individual 
reflections:
$\hat{V}(x_{1},x_{2},x_{3}) = \hat{V}(x_{1}, x_{2},-x_{3})$
etc. This implies that in the 
Fourier transform $\hat{v}(\vec{k})$: ~
$\hat{v}(k_{1},k_{2},k_{3}) = \hat{v}(k_{1},k_{2},-k_{3})$.
Thus, taking only the even part in $k_{3}$
of $\cos(k_{2}+ i \lambda k_{3})$ we will only
be left with the terms $\cos k_{2} \cosh k_{3}$.
Now, $\cosh k_{3}$ is continuous and periodic 
when the Brillouin Zone is 
chosen as $ -\pi < k_{3} \le \pi$. 
No Gibbs phenomenon occurs. The left and 
right derivatives are still discontinuous 
at the zone edges.
\bibitem{z-z}
By replacing the scalar product of Eqn.(\ref{te})
by a non-rotationally invariant product of 
the form 
\begin{eqnarray}
\vec{S}(\vec{x}) \circ \vec{S}(\vec{y}) = \sum_{\alpha =1}^{n}
[S_{\alpha} (\vec{x})]^{z} [S_{\alpha}(\vec{y})]^{z}
\end{eqnarray}
with a power $z  \neq 1$, the rotationally 
invariant Bessel functions (with prefactors) 
of Eqn.(\ref{Bes}) are replaced by 
\begin{eqnarray}
F(\vec{\eta}) = \sum_{\alpha=1}^{n} \int d \hat{\Omega} ~
\eta_{\alpha} \exp[i \eta_{\alpha} \hat{\Omega}_{\alpha}^{z}]
\end{eqnarray}
with  $\hat{\Omega}_{\alpha}$ denoting the 
$\alpha$ component of the n-dimensional 
unit vector $\hat{\Omega}$. The rotational 
invariance of the Bessel function terms in Eqn.(\ref{Bes})
is replaced by the point group symmetry 
of $F(\vec{\eta})$ (that of the 
n-dimensional 
hypercube).

By repeating the steps leading to Eqn.(\ref{super-comp}), 
we find that the partition function of 
\begin{eqnarray}
H = - \sum_{\langle i j \rangle~ along~(1,1)} \hat{V}_{ij} [S_{i}^{(1)}]^{z}
[S_{j}^{(1)}]^{z} \nonumber
 \\
- \sum_{\langle i j \rangle~ along~(1,-1)} \hat{V}_{ij}
[S_{i}^{(2)}]^{z} [S_{j}^{(2)}]^{z} \label{super-dec-z}
\end{eqnarray}
is identical to that of a one dimensional system with
\begin{eqnarray}
H = - \sum_{\langle i j \rangle} \hat{V}_{ij} \vec{S}_{i} \circ \vec{S}_{j}.
\end{eqnarray}
\bibitem{U1} For the $U(1)$ case, the real action
\begin{eqnarray}
S_{R} = -  K \mbox{Re} \{\sum_{\Box} (UUUU)\}. 
\end{eqnarray}
As usual, in the non-Abelian case, the real part 
of trace of 
the plaquette 
product must be 
taken.
\bibitem{Triangular}
It might be easier to compute the terms
in the $\beta$ expansion for a
three dimensional Ising nearest neighbor
ferromagnet by going to 
a triangular lattice and
deleting all
bad ``constructive interference''
loop integrals. 
A line threading the triangular lattice will
connect a spin with spins of distance
$Range=1,$ $n,$ and $n+1$ 
along the line. 
``Bad interference terms'' can result
from loop integrals
of the form $(\exp[ik]  \exp[ink] \exp[-i(n+1)k])^{p}$.
This is the only way in which loop integrals
can conspire to give nonzero results
when $n \rightarrow \infty$.
Thus, trivially, the 
high temperature expansion
for $d=3$ cubic ferromagnet
is equivalent
to that of the triangular
ferromagnet up to 
order $\beta^{3-1}$.
Similarly, if we set 
$s=n-1$ we obtain a
2-Dimensional lattice
model in which each 
spin interacts with
its four nearest neighbors
and four diagonal neighbors.
\bibitem{err}

In the continuum limit,
this corresponds to the kernel
\begin{eqnarray}
\hat{V}(x) = - \frac{\pi}{\Lambda} \sum_{l=1}^{d} \delta(|x-c_{l}|)
\end{eqnarray}
or, in a lattice approximation ($-\pi < k_{1} \le \pi$), to 
\begin{eqnarray}
\hat{V}(x_{1}) = - 2  \sum_{l=1}^{d} 
[(c_{l}+x_{1}) \sin \pi(c_{l}-x_{1}) \nonumber
\\
+ (c_{l}-x_{1})] \sin
\pi(x_{1}+c_{l})]\nonumber
\\ \times [(c_{l}-x_{1})(c_{l}+x_{1})]^{-1}, 
\label{redu}
\end{eqnarray}
(i.e. a scenario in which each
spin is effectively composed of $d$ shifted ``Coulombic sources''
(more precisely, each spin is 
composed of $d$ pairs of ``charges'' 
generating shifted sinc potentials)).
\bibitem{EQU} The astute reader will note that the partition
function of the $d-$ dimensional nearest 
neighbor Ising model on a lattice (with $ -\pi < k_{l} \le \pi$) 
is equivalent to that of the $d-$ dimensional nearest neighbor 
Ising model in the continuum (with the same kernel 
$\hat{v}(\vec{k})$ but with unrestricted  ($- \Lambda < k \le \Lambda$))
\bibitem{perma} That permutational symmetry cannot, for finite $n$, hold to 
all orders in $\beta$ immediately
follows from the fact that 
the normalization constraints
of the $O(n)$ spins are
not permutationally symmetric.
If
\end{thebibliography}
\end{document}